\documentclass[fleqn,usenatbib]{mnras}

\usepackage[utf8]{inputenc}
\usepackage{newtxtext,newtxmath}
\usepackage[T1]{fontenc}
\usepackage{ae,aecompl}
\usepackage[mathscr]{euscript}

\usepackage{float}
\usepackage{graphicx}
\usepackage{subfig}
\usepackage{amsmath}
\graphicspath{{./Images/}}
\usepackage[dvipsnames]{xcolor}
\usepackage{dblfloatfix}

\newcommand{\tenhz}{{\rm 10Hz}}

\newcommand{\beq}{\begin{equation}}
\newcommand{\eeq}{\end{equation}}
\newcommand{\beqn}{\begin{eqnarray}}
\newcommand{\eeqn}{\end{eqnarray}}

\defcitealias{Fabj24}{FS24}
\defcitealias{Sirko03}{SG}

\definecolor{cerulean}{rgb}{0.0, 0.48, 0.65}

\definecolor{navy}{rgb}{0.2, 0.0, 1.0}

\definecolor{jungle}{rgb}{0.0, 0.5, 0.0}

\usepackage{soul}

\title[Spin-Orbit Misalignment in AGNs]
{Spin-Orbit Misalignments of Eccentric Black Hole Mergers in AGN Disks} 
\author[G. Fabj et al.]{
Gaia Fabj$^{1,2}$\thanks{E-mail: gaia.fabj@nbi.ku.dk},
Christopher Tiede$^{1}$,
Connar Rowan$^{1}$,
Martin Pessah$^{1}$,
Johan Samsing$^{1,2}$
\\
$^{1}$Niels Bohr International Academy, Niels Bohr Institute, Blegdamsvej 17, DK-2100 Copenhagen Ø, Denmark\\
$^{2}$Center of Gravity, Niels Bohr Institute, Blegdamsvej 17, 2100 Copenhagen, Denmark
}

\date{Accepted XXX. Received YYY; in original form ZZZ}
\pubyear{2025}

\begin{document}
\label{firstpage}
\pagerange{\pageref{firstpage}--\pageref{lastpage}}
\maketitle

\begin{abstract}
The disks of Active Galactic Nuclei (AGNs) provide a natural environment where stellar mass-black holes (BHs) can dynamically pair, undergo repeated interactions, and eventually merge.
It is commonly assumed that gas accretion will both efficiently spin-up disk-embedded black holes and align the orbits of embedded binaries with the disk plane, leading to mergers with preferentially positive effective spin parameters ($\chi_{\rm eff}$). 
Such predictions have motivated the use of $\chi_{\rm eff}$ as a diagnostic for identifying candidate AGN-embedded mergers in the LIGO-Virgo-KAGRA gravitational wave catalog.
In this work, we perform post-Newtonian N-body simulations of nearly planar binary-single encounters and apply an empirically motivated gas-driven alignment prescription to characterize the expected $\chi_{\rm eff}-$eccentricity correlations of AGN-embedded mergers.
By comparing the alignment and gravitational wave inspiral timescales, we identify the regions of parameter space, across both disk location and binary properties, where full disk-spin-orbit alignment is effective and where it is not. 
We find that quasi-circular binaries typically align by the time they merge, supporting the standard picture of spin-orbit aligned orientations. By contrast, eccentric binaries (with in-band eccentricity $e_{\rm 10Hz}\gtrsim0.1$) typically inspiral too quickly 
for gas torques to act, preserving the post-encounter spin–orbit misalignments and yielding more isotropic $\chi_{\rm eff}$ distributions when disk densities and torque efficiencies are modest. 
This interplay naturally establishes a correlation between binary eccentricity and $\chi_{\rm eff}$ in AGN disks, highlighting a new key observable of the AGN channel and a potential explanation for massive events such as GW190521 and GW231123.
\end{abstract}

\begin{keywords}
stars: kinematics and dynamics -- galaxies: active -- accretion, accretion discs -- galaxies: nuclei -- stars: black holes -- gravitational waves
\end{keywords}

\section{Introduction}

A decade after the first gravitational wave discovery \citep{FirstGW}, the LIGO/Virgo/KAGRA (LVK) 
catalog has grown rapidly, revealing a diverse population of binary black hole (BBH) mergers. Yet, the dominant pathway that produces these systems remains uncertain, with different proposed scenarios.
Competing theories include isolated binary evolution in galactic fields \citep{2012ApJ...759...52D, 2013ApJ...779...72D, 2015ApJ...806..263D, Kinugawa2014MNRAS,2016ApJ...819..108B,
2016Natur.534..512B, 2017ApJ...836...39S, 2017ApJ...845..173M, 2018ApJ...863....7R, 2018ApJ...862L...3S}, primordial black holes
\citep[e.g.][]{2016PhRvL.117f1101S,
2016PhRvL.116t1301B,
2017PDU....15..142C}, 
dynamical interactions encounters in globular clusters (GCs)
\citep{2000ApJ...528L..17P, Lee:2010in,
2010MNRAS.402..371B, 2013MNRAS.435.1358T, 2014MNRAS.440.2714B,
2015PhRvL.115e1101R, 2016PhRvD..93h4029R, 2016ApJ...824L...8R,
2016ApJ...824L...8R, 2017MNRAS.464L..36A, 2017MNRAS.469.4665P, Samsing18, 2018MNRAS.481.4775D, 2018MNRAS.tmp.2223S, 2019arXiv190711231S,2022ApJ...931..149R},
mergers occurring in the nuclear star clusters (NSCs) of Galactic Nuclei 
\citep{2009MNRAS.395.2127O, 2015MNRAS.448..754H,
2016ApJ...828...77V, 2016ApJ...831..187A, 2016MNRAS.460.3494S, 2017arXiv170609896H, 2018ApJ...865....2H} and in
the disks of Active Galactic Nuclei (AGNs)
\citep{2017ApJ...835..165B,  2017MNRAS.464..946S, 2017arXiv170207818M, Tagawa20a,
Samsing22,
2023MNRAS.521..866R,
2023ApJ...944L..42L, 
Delaurentiis23,
2024MNRAS.533.1766W,2023MNRAS.521..866R,Rowan23b,
Rowan2024_rates,Rowan24a,Rowan2025_triples,
Fabj24,
Wang2025_triples,
Wang2025_triplesII, Delfavero2025}. Each channel is expected to leave characteristic imprints on the properties of the binary that in theory can be used to distinguish between dominant merger origins. 
The primary properties of relevance are the merger's mass ratios, orbital eccentricity, and spin orientations. 
Field binaries formed in isolation are expected to mainly produce approximately equal-mass systems merging on circular orbits, due to their evolution through mass transfer and common envelope phases \citep[e.g.][]{2012ApJ...759...52D}. In addition, their individual spin components are expected to be predominantly aligned with the direction of the orbital angular momentum \citep[e.g.][]{2000ApJ...541..319K, Tagawa20, Avi2022}, resulting in a distribution of positive values for the effective spin parameter $\chi_{\rm eff}$. On the other hand, binaries formed through dynamical interactions and strong few-body encounters inside dynamical environments such as GCs and NSCs, are expected to merge with a variety of mass ratios and retain significant eccentricity in the LVK observable band \citep[e.g.][]{2016PhRvD..94h4013C,2009MNRAS.395.2127O,Samsing14,Samsing18}. In addition, dynamical mergers are expected to have an isotropic distribution of spin orientations, producing $\chi_{\rm eff}$ values clustered around zero with both positive and negative tails \citep[e.g.][]{2016ApJ...832L...2R}. 
However, the analysis of detected sources indicates that neither isolated binary evolution nor dynamical formation within globular or nuclear star clusters alone can fully explain the characteristics of the black hole population observed so far \citep{2021ApJ...910..152Z}. 
This discrepancy motivates the search for additional pathways capable of reproducing the observed diversity.
One promising example is the AGN channel, which combines aspects of both isolated and dynamical evolution within a gas-rich environment.

AGN disks represent a more complex, and yet very promising, scenario that combines elements of both the isolated and dynamical channel of binary formation and merger, with the addition of environmental effects due to e.g. the gaseous disk \citep{ 2023ApJ...944L..42L, Delaurentiis23,2023MNRAS.521..866R,  Rowan23b,Rowan24a, 2024MNRAS.533.1766W, Whitehead2025_adiabtic} and tides from the central supermassive black hole (SMBH) \citep{Trani19,2024A&A...683A.135T,2024ApJ...964...43R,Fabj24}. In an active galaxy, black holes in the NSC will be captured into the disk as a result of gas-drag processes \citep{Fabj20,Nasim22, Generozov22,2024MNRAS.528.4958W,Hydrocap1,Hydrocap2}.
Once fully embedded, they can migrate towards the SMBH, grow to large masses, and undergo repeated encounters (scatterings) potentially yielding multi-generation and hierarchical mergers. 

Orbits of stars residing in the NSC can also be aligned with the plane of the disk and undergo modified stellar evolution due to gas accretion \citep{Jermyn21,Cantiello21,2021ApJ...916...48D,2022ApJ...929..133J,2023ApJ...946...56D,Fabj25, 2025ApJ...979..245D}, producing a large fraction of massive stars and black holes.
Gas not only tends to align inclined stellar objects but it will also align spin-orbit tilted binaries \citep[e.g.][]{2004ApJ...602..388T, Tagawa20,2024MNRAS.531.3479M}, which can originate from dynamical interactions. This implies that a population of AGN-channel mergers would be biased toward positive values of $\chi_{\rm eff}$ \citep[e.g.][]{Tagawa20a,Mcfacts1, Mcfacts2}. 

Two events stand out as possible AGN-channel sources. The first is GW190521, one of the most massive BBH mergers detected to date, with a total mass of about $150 \, \rm M_\odot$ \citep{GW19a}. Its inferred properties represent a challenge to the standard isolated binary and dynamical formation scenarios because the component masses lie in or above the pair-instability mass gap \citep[e.g.][]{2020ApJ...904L..26F,2020ApJ...891L..31F,2021ApJ...907L...9N}, the waveform is consistent with a possible eccentric merger of $e\ge 0.1$ at 10 Hz \citep[e.g.][]{2020ApJ...903L...5R,2022ApJ...940..171R,2022NatAs...6..344G}, and the measured spin–orbit tilt suggests a large misalignment between the component spins and the orbital plane \citep{GW19a}.
The combination of features have motivated the idea that GW190521 could have formed in an AGN disk, where hierarchical mergers \citep{Tagawa21} and accretion \citep{2012MNRAS.425..460M,2022ApJ...928..191G} can grow massive black holes, while few-body encounters can generate eccentric orbits \citep[e.g.][]{Samsing22,Fabj24,Rowan2025_triples}.

The second source of interest is GW231123, the most massive BBH merger up to date, $\sim 200 \, \rm M_{\odot}$, with evidence for individual highly spinning components \citep{2025arXiv250708219T}. 
Such rapid spins are difficult to explain in purely dynamical environments, where black holes typically retain low natal spins or experience spin-down during hierarchical mergers \citep[e.g.][]{2021NatAs...5..749G,2019PhRvD.100d3027R}, but could naturally arise in AGN disks if sustained accretion efficiently spins up black holes \citep{2025arXiv250813412D,2025arXiv250808558B}.

Furthermore, the geometry of dynamical interactions in AGN disks plays an essential role in producing a high number of mergers. In the coplanar 2D-disk limit, in contrast with 3D clusters, all objects share the same orbital plane, maximizing angular momentum exchange, the efficiency of hardening, and the production of eccentric mergers \citep{Samsing22}. 
At the same time, introducing even a small inclination angle shifts the dynamics toward the three-dimensional scattering regime, suppressing the probability of eccentric mergers by one to two orders of magnitude once the inclination exceeds about a degree \citep{Samsing22, Fabj24}.
The occurrence of non-coplanar encounters is quite plausible, as disk turbulence can inhibit full alignment and maintain small but significant orbital inclinations \citep[e.g.][]{Chen2023_chaotic_acc,2025arXiv250602173T}.
This argument motivates the key question addressed in this work: can dynamically assembled binaries in AGN disks, particularly those that merge with significant eccentricity, retain spin–orbit tilts despite the presence of gas?
We provide a first answer to the question, investigating the interplay between gas-driven alignment and gravitational wave inspiral while focusing on how this competition shapes the final inclination and spin properties of merging binaries.

In this work, we combine 3D post-Newtonian 
N-body simulations of binary–single encounters with an analytic prescription for gas-driven alignment. The paper is organized as follows: in Sec.~\ref{sec:models} we describe the AGN disk model, scattering setup, and alignment prescription. The simulations establish the inclination distribution of dynamically formed binaries at formation, while the alignment model allows us to compare the characteristic timescales of gas accretion and gravitational wave inspiral time across a range of disk properties. In Sec.\ref{sec:timescales}, we define the GW and alignment timescales, while outlining their main dependencies.
In Sec.~\ref{sec:results}, we show our results with two different analyses. We first map how alignment efficiency varies with eccentricity, semi-major axis, and disk location, highlighting the parameter space where binaries are expected to retain their spin–orbit tilts up to merger. We then present the inclination distributions for the N-body simulations and show their evolution under different alignment efficiencies, illustrating the contrast in distribution between the circular and eccentric mergers.
In Sec.~\ref{sec:caveats} we discuss caveats and limitations of our framework. In Sec.~\ref{sec:conclusion}, we summarize our conclusions and observational implications for the AGN channel, particularly the expected $\chi_{\rm eff}$ distributions and the possibility for eccentric, gas-assisted mergers in AGN disks. We conclude by providing an outline and directions for future work.

\begin{figure*}
    \centering
    \includegraphics[width=\linewidth]{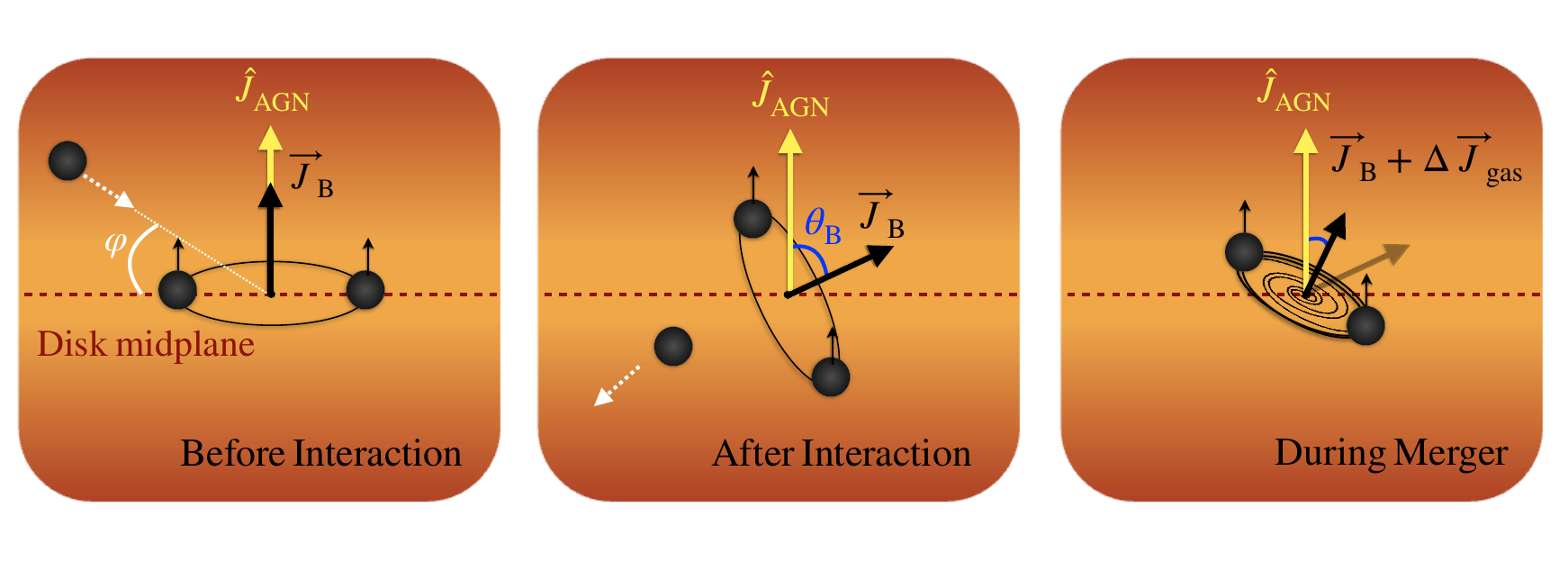}
    \caption{Sketch of out-of plane of a typical BH binary-single interaction in an accretion disk. The small black arrows represent the black hole individual spins, which due to prior disk interaction, are aligned with the angular momentum of the disk $\hat{J}_{\rm AGN}$ and remain aligned throughout the process.
    Left panel: configuration of the three-body system before the encounter. The single approaches at some angle $\varphi$ the binary with its orbital angular momentum $\vec{J}_B$ aligned with the one of the disk  $\hat{J}_{\rm AGN}$. Middle panel: the interaction with the single causes a kick tilting $\vec{J}_B$ at an angle $\theta_B$. Right panel: while the binary is merging due to GW radiation, it experiences gas accretion adding momentum $\Delta \vec{J}_{\rm gas}$ onto $\vec{J}_B$, thus realigning the binary with the disk. The figure is re-adapted from \citet{Samsing22}.
    }
    \label{fig:sketch}
\end{figure*}

\section{Models} \label{sec:models}

First, we describe our choice of AGN disk model, for which we adopt the dense, marginally stable \citet{Sirko03} (hereafter \citetalias{Sirko03}) disk as a reference framework for evaluating disk-dependent quantities.
We then present our post-Newtonian N-body simulations of three-body scatterings, designed to capture how binary–single encounters inside the disk generate spin–orbit tilts.
Next, we introduce the schematic picture of spin–orbit tilt evolution for dynamically assembled binaries in disks, emphasizing the interplay between misalignment through scattering and subsequent re-alignment through gas accretion.
Finally, we outline our alignment prescription, in which the angular momentum evolution of the binary is modeled through gas accretion in different disk regimes, allowing us to quantify how efficiently binaries can re-align compared to their gravitational wave inspiral timescales.

\subsection{AGN Disk Model}

We consider the \citetalias{Sirko03} disk model
for a range of SMBH masses ($10^6$–$10^9 \, \rm M_{\odot}$), computed with the \texttt{pAGN} package \citep{pAGN}. 
The \citetalias{Sirko03} model extends the classical \citet{Shakura73} thin-disk framework to parsec scales while preventing fragmentation by enforcing marginal stability ($Q \gtrsim 1$) through additional heating. This prescription produces disks that are hotter and denser at large radii than standard $\alpha$-disks, with opacity transitions accounted for using tabulated opacity tables \citep{Semenov03,Badnell05}.
Compared to other prescriptions, such as the \citet{Thompson05} model, the \citetalias{Sirko03} disks are systematically denser, particularly in their inner regions, between $10^2-10^4 \, \rm R_s$, where $R_s$ is the Schwarzschild radius. This makes \citetalias{Sirko03} a natural choice to highlight the regime in which alignment can take place most efficiently when merging binaries are embedded in the densest parts of AGN disks.
In environments where embedded compact objects provide significant local heating, the disk may naturally approach a marginally stable configuration as well \citep[e.g.][]{2025MNRAS.537.3396E}, further motivating the focus on the \citetalias{Sirko03} prescription.
As standard parameters in \texttt{pAGN}, we set the viscosity to $\alpha = 0.01$ and the Eddington fraction to $l_{\rm E} = 0.5$. 

\subsection{Three-Body Scatterings}\label{sec:scatterings}

We carry out post-Newtonian (PN) N-body simulations of black hole (BH) binary–single interactions (three-body scatterings) adopting the same methods introduced in \citet{Fabj24} (referred to \citetalias{Fabj24} from here on). 
The scattering experiments account for the tidal field of the central SMBH in the equations of motion but neglect the influence of gas during the encounter. The contribution  from the SMBH implemented in \citetalias{Fabj24} extends the analysis of \citet{Samsing22} on the probability of dynamically forming an eccentric merger based on the constrained geometry in AGN disks. Besides \citetalias{Fabj24}, the overall code has also been extensively tested in previous studies \citep{Samsing14,Samsing17,Samsing18,Samsing22}, and we refer the reader to Sec.~3 of \citetalias{Fabj24} for further details on the numerical implementation and simulation setup.

As the specific scope of this paper is to study the binary spin–orbit tilt distribution in a disk setup, we focus on a three-dimensional encounter geometry. 
The single incoming BH approaches at a small inclination angle a binary located in the disk mid-plane ($\varphi = 0.1^{\circ}$, see first panel of Fig.~\ref{fig:sketch}). This represents a likely scenario in AGNs, where the three BHs are initially embedded in the accretion disk, and the vertical offset of the incoming object does not exceed the disk opening angle, set by its  characteristic vertical thickness. 
While our chosen $\varphi$ is small, such out-of-plane perturbations are particularly relevant, as even a low inclination can break the symmetry of coplanar scatterings.
In fact, the likelihood of producing eccentric mergers is highly sensitive to $\varphi$, with rates dropping by nearly an order of magnitude for
$\varphi= 1^{\circ}$ compared to the coplanar case \citep{Samsing22,2024A&A...683A.135T, Fabj24}. 
Recent hydrodynamical simulations \citep[e.g][]{Hydrocap2,Hydrocap1} have shown that the single objects on lowly inclined orbits ($\varphi<3^
\circ$) rapidly embed themselves in the AGN disk, within order $\sim$10 orbits around the SMBH, making the assumption of a relatively low $\varphi$ justified.
We model the interaction such that the inclined single BH is placed at the Hill radius of the triple system with respect to the SMBH, with a velocity vector equivalent to the coplanar case, but with zero velocity in the $z$ direction at the start of the simulation.
We adopt an equal-mass three-body system with component masses of $15 M_{\odot}$, and with mass ratios relative to the SMBH in the range $10^{-5}$–$10^{-8}$. The binary is placed in the disk mid-plane and is initially circular with a semi-major axis of 1 AU.

\subsection{Spin-Orbit Tilt Evolution of Dynamical Mergers in the Disk}\label{sec:tilt_methods}

Fig.~\ref{fig:sketch} shows a schematic representation of the evolution of the binary 
angular momentum for a BBH merger resulting from a three-body interaction embedded 
in an AGN disk. 
The left panel illustrates the initial configuration of the triple system: the BBH lies 
in the disk mid-plane (possibly having formed through gas-assisted dynamical formation), 
with the incoming single approaching at inclination $\varphi$ relative to the disk plane. 
Before the interaction, the binary angular momentum $\vec{J}_{\rm B}$ is assumed to be aligned with 
the overall direction of the disk angular momentum $\hat{J}_{\rm AGN}$. 
The individual BH spins are also assumed to be aligned, as accretion torques are expected 
to bring them into alignment with the AGN disk angular momentum vector on short 
timescales \citep[e.g.][]{Tagawa20,Avi2022, Santini2023,2024arXiv240905614M}. 
After the encounter (middle panel), the binary receives a dynamical kick, producing a 
misaligned spin–orbit tilt $\theta_B$ relative to $\hat{J}_{\rm AGN}$. 
While $\theta_B$ characterizes the orientation of the binary orbital angular momentum, its cosine directly enters the 
definition of the effective spin parameter $\chi_{\rm eff}$ under the assumption that 
individual black-hole spins are aligned with the AGN disk \footnote{
This assumption neglects possible misalignment of individual black-hole spins due to turbulent accretion via mini-disks or due to supersonic motion of the binary relative to the AGN disk, which can reorient mini-disks \citep[e.g.,][]{2022ApJ...928L...1L,2022ApJ...939L..23C}.
}
. 
Thus, the distribution of $\theta_B$ maps directly into the expected distribution 
of $\chi_{\rm eff}$. 
As the tilted binary evolves toward merger (right panel), it accretes gas, and some angular momentum $\Delta \vec{J}_{\rm gas}$ is added to $\vec{J}_{\rm B}$, 
driving the system back towards alignment.

A key question is whether the binary has time to fully realign before merging. 
Both the alignment timescale and the gravitational wave merger timescale can be 
very short depending on the binary’s compactness and eccentricity. 
For instance, eccentric systems may inspiral extremely rapidly, and at the 
same time dense gas environments can drive alignment on comparably short 
timescales. Which process occurs faster is non-trivial and to predict it requires 
a detailed comparison across parameter space (i.e. binary and disk properties).

\subsection{Alignment prescription}\label{sec:prescription}
We define $\dot{J}_z$ as the rate of change for the binary angular momentum component in the direction of the AGN disk angular momentum $\hat{J}_{\rm AGN} = \hat{J}_z$. 
Following the prescription of \citet{Tagawa20}, we write
\begin{equation}\label{eq:jdot}
\dot{J}_z = a_{\rm bin} \, v_{\rm bin} \, \dot{M} \, f_{\rm rot},
\end{equation}
where $a_{\rm bin}$ is the binary semi-major axis, $v_{\rm bin}$ its orbital velocity, $\dot{M}$ the mass accretion rate defined later, and $f_{\rm rot}$ a dimensionless parameter characterizing the typical angular momentum exchange of a gas parcel. $\dot{J}_z$ is constant during the alignment process, allowing us to define a characteristic alignment timescale in terms of the binary and disk properties, later expressed through Eq.~\eqref{eq:tau}. Similar to the method used in \citet{Fabj25} to map the different regimes of stellar gas accretion in AGNs, we set a control radius and consider different regimes inside the disk
to properly describe $\dot{M}$. We compare the size of the Bondi radius 
\begin{equation}
R_{B} \equiv \frac{2 G \,m_{\rm bin}}{c_s^2},
\end{equation} 
and the Hill radius
\begin{equation}
R_H \equiv \left(\frac{Gm_{\rm bin}}{3\Omega_{\bullet}^2}\right)^{1/3},
\end{equation}
where $m_{\rm bin}$ is the mass of the binary, $c_s$ is the local sound speed in the disk, and $\Omega_{\bullet}$ is the binary orbital frequency around the SMBH.

We note that using the total sound speed implicitly assumes optically thick, slow-diffusion conditions under which the radiation field remains coupled to the gas \citep{2024ApJ...974..106C}. 
In \citetalias{Sirko03} disks, radiation pressure is important mainly in the innermost regions (typically $\leq 10^3R_g$, e.g \citealt{GoodmanTan2004,pAGN}), encompassing only a small fraction of embedded stellar-mass black holes \citep[e.g.][]{2017MNRAS.464..946S} and where the optical depth is high.  The fast-diffusion regime, in which radiation pressure decouples from the gas 
($\tau \lesssim c/c_s$), is not modeled here, as it mainly occurs in the outer disk 
where gas pressure dominates and radiation pressure does not significantly influence the Bondi radius.

When $R_B<R_H$, we refer to the binary as being in an accretion-dominated (\textbf{AD}) regime of the disk. On the other hand, we refer to the binary as being in a tidally limited (\textbf{TL}) regime when $R_H<R_B$. 
It follows that the mass accretion rate onto the binary can be approximated as
\begin{equation}\label{eq:mdot}
\dot{M} = \rho_{\rm g} c_s\left(\min\left\{R_B,R_H \right\}\right)^2,
\end{equation}
where $\rho_{\rm g}$ is the disk density. The accretion rate is set by the smaller of the two characteristic radii, since gas accretion is limited by the more restrictive scale.
In order to properly reflect that accretion begins at the boundary set by the minimum value between $R_B$ and $R_H$,
we further restrict our analysis to binaries that are well embedded within their Hill sphere. This ensures we do not overestimate the accretion rate for binaries whose semi-major axis would otherwise extend close to the Hill radius.

When $\dot{M}$ is in the accretion dominated regime we model it as $ \dot{M}_{\rm AD} = 0.08 \, \dot{M}$, corresponding with recent hydrodynamical simulations of inclined \citep{Dittmann24} and eccentric \citep{2024ApJ...970..107C} binaries. 
However, we note that this generally corresponds to accretion rates far exceeding the binary's Eddington limit (see Section~\ref{sec:caveats}),
so we do not explicitly assume that the inflowing gas is deposited into the black holes.
Instead, similar to the analysis of \citet{Tagawa20}, we quantify the angular momentum exchange per inflowing gas parcel phenomenologically through the parameter $f_{\rm rot}$ where for $f_{\rm rot} = 1$ the transferred angular momentum corresponds to that of a circular orbit at $a_{\rm bin}$. 
A lower value of $f_{\rm rot}$ (e.g., 0.1) 
corresponds to less torque applied to the binary, for say, less rotationally supported gas. 
On the other hand, a very high efficiency (e.g., $f_{\rm rot}$ = 10), translates to ten times more angular momentum transport per unit mass than the circular Keplerian case. 

The hydrodynamical simulations in \citet{Dittmann24} investigate the evolution of initially circular binaries placed at a small inclination (both highly prograde and retrograde) with respect to the disk mid-plane. By taking the ratio of the reported rate of inclination change to the initial tilt angle, we infer a characteristic alignment timescale and compare it with Eq.~\eqref{eq:jdot}. 
We find that the \citet{Dittmann24} results correspond to a value of $f_{\rm rot}=0.1$ in our model, which we identify as our fiducial case. However, since our prescription is phenomenological and neglects some physical processes (see Sec.~\ref{sec:caveats}),we also include results for $f_{\rm rot}=1,10$  to account for the possibility that the chosen prescription underestimates the alignment efficiency. In this sense, $f_{\rm rot}$ can be interpreted as a simple parameterization of the gas inflow into the binary,
ranging from less efficient to more efficient alignment. 
Further hydrodynamical simulations will be necessary to determine which regime is the most representative of actual AGN disk conditions.

\section{Timescales}\label{sec:timescales}

\subsection{Gravitational Wave Timescale}

In order to evaluate the gravitational wave timescale $t_{\rm gw}$ we assume that the orbital evolution follows Peters' equations \citep{1964PhRv..136.1224P}, using them to relate orbital separation, frequency, and eccentricity. The timescale $t_{\rm gw}$ for an equal-mass binary can be approximated as \citep{1964PhRv..136.1224P, Samsing22}:

\begin{equation}\label{eq:tgw}
t_{\rm gw} \approx \frac{5c^5}{512G^3} \frac{a_{0}^4}{m_{\rm BH}^3} \big( 1-e_0^2 \big)^{7/2},
\end{equation}
where $a_0$ and $e_0$ are the binary semi-major axis and eccentricity at the time of formation (i.e. immediately after the binary–single encounter ends and the dynamically assembled binary begins inspiraling). The steep dependence on eccentricity in Eq.~\eqref{eq:tgw} makes the merger timescale extremely sensitive to this quantity, as eccentric binaries merge rapidly. This occurs because in the GW-driven regime each pericenter passage extracts significant orbital energy, even as the binary circularizes.

Though we evaluate both the alignment and coalescence timescales starting from the binary properties at formation, 
throughout this paper we express our results in terms of the eccentricity when the GW signal enters the LVK band, 
defined at $f_{\rm gw} = 10 \, \rm Hz$. 
The GW peak frequency is the characteristic frequency where most of the gravitational wave power is radiated 
\citep{DJD18}.  
For three-body mergers, where the binary begins inspiraling while the third body is still bound, 
the systems are typically formed with very high eccentricities \citep[e.g.][]{Samsing18,Samsing22}. 
In such cases, gravitational wave emission is strongly concentrated near pericenter, 
and the corresponding pericenter distance  
\begin{equation}
r_p \approx \left(\frac{G \, m_{\rm bin}}{{\pi}^2 f_{\rm gw}^2}\right)^{1/3}
\label{eq:rf}
\end{equation}
sets the GW peak frequency.
From each scattering experiment obtained through the post-newtonian (PN) $N$-body simulations we obtain the binary’s orbital elements at the end of the three-body encounter, namely the semi-major axis $a_{\rm out}$ and eccentricity $e_{\rm out}$. From these we compute the constant $c_{\rm out}$ from \citep{Peters64}, which represents the conserved combination of orbital parameters that defines the binary evolutionary track under GW emission: 

\begin{equation}\label{eq:c0}
c_{\rm out} = \frac{a_{\rm out}(1-e_{\rm \rm out}^2)}{e_{\rm \rm out}^{12/19}\left[1+\frac{121}{304} e_{\rm out}^2\right]^{870/2299}}.
\end{equation}

A large fraction of the mergers resulting from the binary-single encounters yield binaries whose GW peak frequency at formation already exceeds 10 Hz. For the remaining systems with $f_{\rm gw} < 10 \, \rm Hz$ at formation, 
we evolve their eccentricity forward in time using Peters' equations. 
Combining Eq.~\eqref{eq:c0} with the relation between $r_p$ and $f_{\rm gw}$, 
we obtain
\begin{equation}\label{eq:rp}
f_{\rm gw}
= 
\frac{\sqrt{G m_{\rm bin}} \,(1+e)^{3/2}}
{\pi \,c_{\rm out}^{3/2}\, e^{18/19}\,
\left(1+\frac{121}{304}e^{2}\right)^{1305/2299}}.
\end{equation}
This expression connects the binary eccentricity $e$ at a given $f_{\rm gw}$ 
to $c_{\rm out}$. For each binary, we solve Eq.~\eqref{eq:rp} numerically at $f_{\rm gw} = 10 \, \rm Hz$, 
which allows us to translate the initial post-encounter orbital elements 
into the eccentricity when the system enters the LVK band.  

To initialize the alignment process at formation, we must determine the orbital parameters at that time, 
rather than only at $f_{\rm gw}=10\,\rm Hz$.  
This requires connecting the observed eccentricity at 10 Hz, $e_\tenhz$, 
to the corresponding eccentricity at formation, $e_0$. We do this using the Peters' invariant \citep[e.g.][]{2017MNRAS.464L..36A,Samsing18}, 
which relates pericenter distance and eccentricity during GW-driven inspiral.  
The relation gives the (initial) pericenter distance $r_0$ for the binary at formation, 

\begin{equation}\label{eq:backP}
r_0 = \left(\frac{Gm_{\rm bin}}{f_{\rm gw}^2 \pi^2}\right)^{1/3} \frac{1}{2} \frac{1+e_\tenhz}{e_\tenhz^{12/19}} \left [ \frac{425}{304} \left( 1 + \frac{121}{304}e_\tenhz^2\right) ^{-1}\right]^{870/2299}.
\end{equation}
Using the identity $r_0 = a_0 (1-e_0)$, we can then solve for $e_0$ from the left-hand side given an estimate of the initial semi-major axis $a_0$, 
which we take to be of the same order as the pre-encounter binary separation.    
The recovered $e_0$ then sets the used initial condition for evaluating alignment and merger timescales.
We note that we have confirmed that despite the very high eccentricities at formation, $a_0$ is always sufficiently large to ensure the validity of Peters' equations.
Finally, 
we note that the actual inspiral time can in principle be shorter than $t_{\rm gw}$ if gas drag or tidal interactions contribute to orbital decay, particularly in dense AGN disk regions. This effect can be especially relevant for e.g. quasi-circular transonic binaries \citep[e.g.][]{2024ApJ...974..216O}. 
However, these effects are not treated in the present model as they would require a full time-dependent solution to the binary orbit evolution, substantially complicating the preliminary framework adopted for this study.
We reserve this implementation for future work.

\subsection{Alignment Timescale}
We model the alignment timescale by tracking how the perpendicular component of the binary’s angular momentum, $J_\perp$, is damped as gas is accreted. 
We assume that the total angular momentum of the binary
\begin{align}\label{eq:jbin}
J &= \sqrt{J_\perp^2 + J_z^2} = \mu \sqrt{G m_{\rm bin} a_0 (1-e_0^2)}
\end{align}
is fixed (with $\mu$ the reduced mass),
such that the torque acts only to rotate the binary angular momentum into alignment with $\hat{J}_{\rm AGN}$.
\hspace{-2pt}\footnote{
This corresponds to fixing the binary semi-major axis.
Of course this will also shrink due to GW radiation and gas torques, but the alignment timescale, Eq.~\eqref{eq:tau}, does not depend on $a_{\rm bin}$, except for any implicit relation through $f_{\rm rot}$. We comment on potential effects of $\dot e_{\rm bin}$ in Sec.~\ref{sec:caveats}.
}
Differentiating the identity $J_z / J = \cos \theta_B$ yields the relation,
\begin{equation}\label{eq:diff}
\sin\theta \, d\theta = - \frac{\dot{J_z}}{J} \, dt ,
\end{equation}
and we define the characteristic alignment timescale as
\begin{equation}\label{eq:tau}
\tau_{\rm align} \equiv \frac{J}{\dot{J}_z},
\end{equation}
where the quantity $\tau_{\rm align}$ carries information of both the binary and disk properties, which are assumed to be constant throughout the process. Integrating Eq.~\eqref{eq:diff} from the initial inclination $\theta_B$ down to perfect alignment ($\theta=0$) gives the total alignment time,
\begin{equation}\label{eq:talign}
t_{\rm align} = \tau_{\rm align}(1-\cos\theta_B).
\end{equation}
This expression highlights the geometric dependence on the initial tilt. For nearly coplanar binaries ($\theta_B << 90^\circ$), alignment occurs rapidly. For $\theta_B = 90^\circ$, the alignment timescale equals $\tau_{\rm align}$. For retrograde binaries alignment takes longer, approaching nearly twice $\tau_{\rm align}$ as $\theta_B \to 180^\circ$ . In our framework, we define "retrograde" as the sources inclined at an angle of $\theta_B \gtrsim 90^\circ$ with respect to $\hat{J}_{\rm AGN}$ such that $J_z$ is initially negative and must be passed through zero to align $J$ with $\hat{J}_{\rm AGN}$), and we assume that they always progress toward prograde orientations, in agreement with the hydrodynamical simulation of \citet{Dittmann24}. We discuss the physical intuition for this in Sec.~\ref{sec:retro}.

Finally, the value of $\dot{J}_z$ depends on the accretion regime. By substituting the expressions for the angular-momentum change in the accretion-dominated and tidally limited regimes, we obtain the corresponding scalings of $\tau_{\rm align}$. The alignment formulations are
\begin{equation}
    \tau_{\rm align,AD} = 
    \frac{c_s^3}{f_{\rm rot} \, \rho_{\rm g} } \,
    \sqrt{\frac{\mu^2 \,(1-e_0^2)}{16 \, \pi^2(G  m_{\rm bin})^4}},
\end{equation}\label{eq:tad}
and
\begin{equation}
    \tau_{\rm align,TL} = 
    \frac{(3\Omega_{\bullet}^2)^{2/3} }{ f_{\rm rot}  \, \rho_{\rm g} \, c_s} 
    \sqrt{\frac{\mu^2 \,(1-e_0^2)}{\pi^2 \,(G  m_{\rm bin})^{4/3}}} .  
\end{equation}
The main distinction between the two regimes lies in their dependence on 
the disk sound speed $c_s$ and the scaling with binary mass. 
In the accretion-dominated case, the alignment timescale depends directly 
on $c_s$, whereas in the tidally-limited regime the dependence enters 
through the orbital frequency, $\Omega_{\bullet} = H/c_s$, with $H$ the disk scale height. 
For the binary mass,
since $R_B \propto m_{\rm bin}$ and $R_H \propto m_{\rm bin}^{1/3}$, 
the alignment efficiency varies accordingly.  
In both regimes, the alignment timescale is linear in the local gas density 
$\rho_{\rm g}$ and the rotation efficiency parameter $f_{\rm rot}$. 
Thus, increasing either the gas density or the alignment efficiency 
accelerates alignment commensurately.

\subsection{Relative Timescales}\label{sec:ratio}
\begin{figure*}
    \centering
    \includegraphics[width=\linewidth]{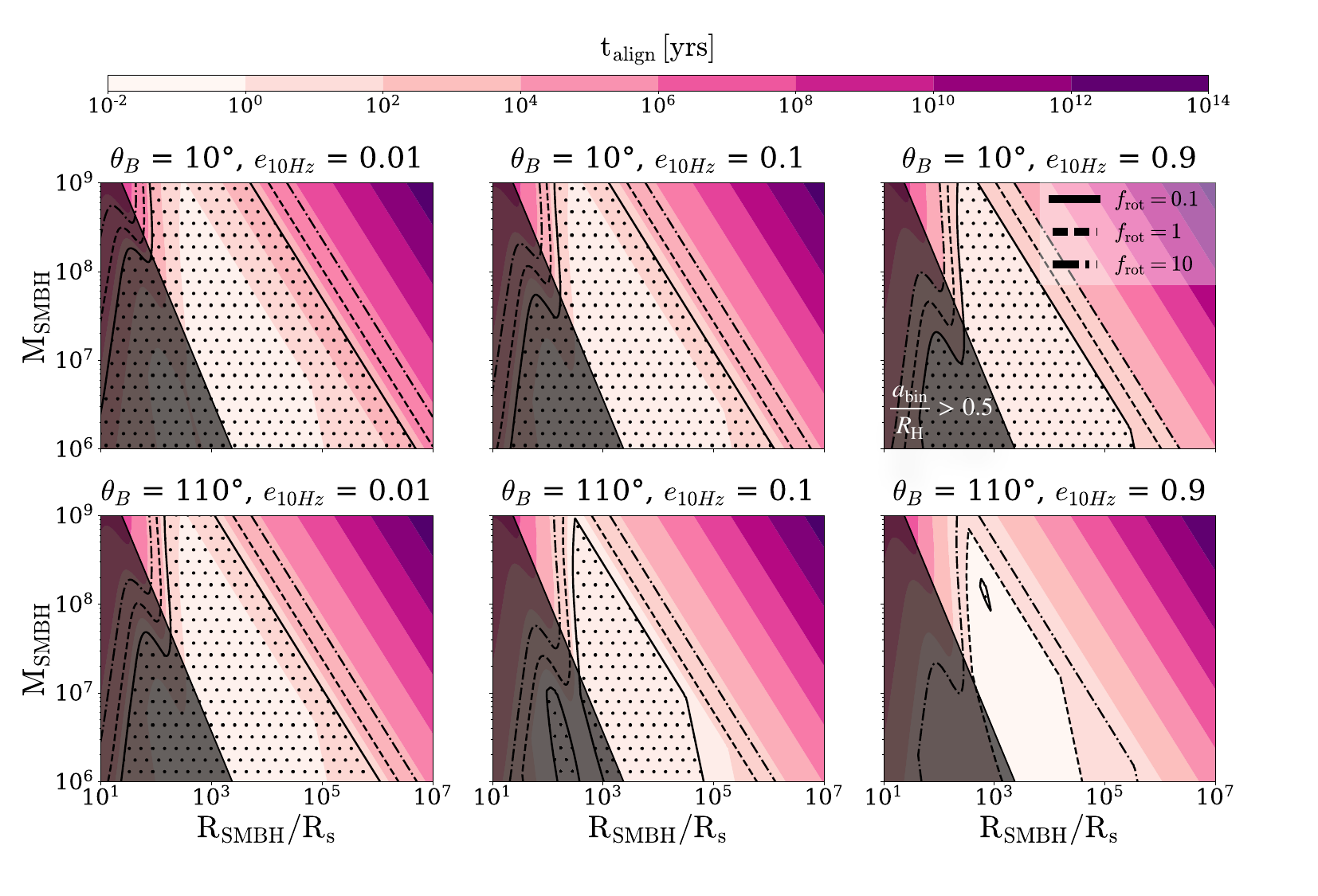}
    \caption{Alignment timescale maps for the \citetalias{Sirko03} disk model, shown as a function of SMBH mass and distance from the SMBH in units of $R_{\rm s}$. We consider a binary with initial semi-major axis $a_0=1$ AU, shown for different eccentricities at 10 Hz ($e_{\rm 10Hz}=0.01, 0.1, 0.9$) and two initial inclinations ($\theta_B=10^\circ, 110^\circ$). The dark shaded region marks where $a_0$ exceeds half of the Hill radius; these cases are excluded as we only consider binaries well embedded within their Hill sphere. Black curves indicate $t_{\rm align}=t_{\rm gw}$, with line style showing alignment efficiency: $f_{\rm rot}=0.1$ (solid), 1 (dashed), and 10 (dot-dashed). The dotted shaded area highlights regions where $t_{\rm align}>t_{\rm gw}$ for the fiducial $f_{\rm rot}=0.1$. Increasing $f_{\rm rot}$ enlarges the parameter space where alignment occurs before merger. According to Eqs.~\ref{eq:ratioad}-\ref{eq:ratiotl}, higher eccentricities shrink the region where $t_{\rm align}>t_{\rm gw}$, particularly at larger inclinations, implying that more binaries merge misaligned.
    } 
    \label{fig:map}
\end{figure*}

Because the primary aim of this work is to compare alignment and 
gravitational wave timescales, we explicitly express their ratio for both 
the AD and TL regimes:

\begin{equation}\label{eq:ratioad}
    \frac{\tau_{\rm align, AD}}{t_{\rm gw}} \approx  \frac{13 G}{ \pi c^5} 
    \frac{c_s^3}{f_{\rm rot} \, \rho_{\rm g}  } 
    \frac{\mu^2 }{a_0^4 \, (1-e_0^2)^3} \, , 
\end{equation}

\begin{equation}\label{eq:ratiotl}
    \frac{\tau_{\rm align, TL}}{t_{\rm gw}} \approx 
    \frac{107 G^{7/3}}{ \pi c^5} \,
    \frac{\Omega^{4/3}}{f_{\rm rot} \, \rho_{\rm g} \, c_s} \,\frac{(\mu \, m_{\rm bin}^{2/3})^2}{a_0^4 (1-e_0^2)^3}.
\end{equation}
These scalings imply that when \(\tau_{\rm align}/t_{\rm gw} > 1\),
GW coalescence outpaces gas-driven alignment, and therefore the binary
reaches merger with a residual spin–orbit misalignment.
Conversely, when \(\tau_{\rm align}/t_{\rm gw} < 1\), alignment proceeds
efficiently and the binary is aligned by the time it merges.
The dependence on $e_0$ plays a large role in determining the relative importance of the two timescales, as later illustrated in Fig.~\ref{fig:map}. 
In particular, higher eccentricities lead to less alignment despite having less angular momenta because $t_{\rm gw}$ shrinks more rapidly than $\tau_{\rm align}$ with growing $e$.
This effect is especially present at very high eccentricities, as even a small difference, such as $e=0.998$ versus $e=0.99$, leads to a dramatic change in the ratio, noting that the timescales are evaluated at formation where binaries are typically highly eccentric. 
We provide scaling relations for the ratio $\tau_{\rm align}/t_{\rm gw}$ in the two 
accretion regimes. For clarity, we restrict the problem to equal-mass binaries and express the 
eccentricity in terms of $e_{\rm 10Hz}$, which can be mapped back to the capture 
eccentricity using Eq.~\eqref{eq:backP}. In the accretion-dominated regime, the ratio takes the form 
\begin{align}
    \frac{\tau_{\rm align, AD}}{t_{\rm gw}} \approx \;&
    1.4 \times 10^2 \,
    \left(\frac{f_{\rm rot}}{0.1}\right)^{-1}
    \left(\frac{\rho_{\rm g}}{10^{-12} \, \rm g/cm^3}\right)^{-1} \left(\frac{c_s}{5 \cdot 10^6  \, \rm cm/s}\right)^{3} \nonumber \\
    & \times 
    \left(\frac{15 \, M_{\odot}}{m_{\rm BH}} \right)^{-2} \bigg(\frac{a_0}{1 \, \rm AU} \bigg)^{-4} \left(\frac{1-e_{\rm 10Hz}^2}{0.75}\right)^{-3}, 
    \nonumber \\
\label{eq:scaling_ad}
\end{align}
while in the tidally-limited regime it becomes
\begin{align}
    \frac{\tau_{\rm align, TL}}{t_{\rm gw}} \approx \;&
    0.4 \, 
     \left(\frac{f_{\rm rot}}{0.1}\right)^{-1}
    \left(\frac{\rho_{\rm g}}{10^{-9} \, \rm g/cm^3}\right)^{-1}
    \left(\frac{\Omega_{\bullet}}
    {1\ {\rm Hz}}\right)^{\frac{4}{3}}
    \nonumber \\
    & \times
    \left(\frac{15 \, M_{\odot}}{m_{\rm BH}} \right)^{-\frac{10}{3}}\bigg(\frac{a_0}{1 \, \rm AU} \bigg)^{-4} \left(\frac{1-e_{\rm 10Hz}^2}{0.75}\right)^{-3}. 
    \nonumber \\
\label{eq:scaling_tl}
\end{align}
The factor of 0.75 corresponds to $(1-0.5^2)$, reflecting the reference case of 
$e_{\rm 10Hz}=0.5$ adopted in these scalings. The main contrast between the two 
regimes shown by the pre-factors (>1 in AD and <1 TL) arises from the disk density: in the TL regime, where densities are much 
higher, and the ratio drops below unity, alignment dominates over merger. 
Here we set $f_{\rm rot}=0.1$ as our fiducial case, but this efficiency parameter 
is equally important in determining whether alignment can outpace inspiral, as larger
values of $f_{\rm rot}$ accelerate realignment proportionally.
At the same time, the steep $a_0^{-4}$ scaling implies that as binaries become more compact (either initially or through gaseous effects) the GW inspiral time becomes faster, effectively limiting the time available for alignment.
Thus, in dense inner regions alignment is generally unavoidable, 
whereas in the outer disk, or for compact and eccentric binaries, misalignment can persist 
until merger.

\section{Results} \label{sec:results}
In this section we present our results on the competition between gas–driven alignment and gravitational wave inspiral timescales in AGN disks. In Sec.~\ref{sec:map} we map alignment timescales across the \citetalias{Sirko03} disk model, identifying the regions where binaries can realign before merging. In Sec.~\ref{sec:spinev} we analyze spin–orbit tilt distributions from PN N-body simulations, both without and with gas alignment, and connect them to the expected distribution $\chi_{\rm eff}$. In Sec.~\ref{sec:uplim} we provide further insights on the upper limit for the alignment angle and finally, in Sec.~\ref{sec:retro} we discuss additional orientation effects such as Bardeen-Peterson torques, showing they are negligible for stellar-mass BBHs in AGNs.

\begin{figure*}
    \centering
    \includegraphics[width=0.8\linewidth]{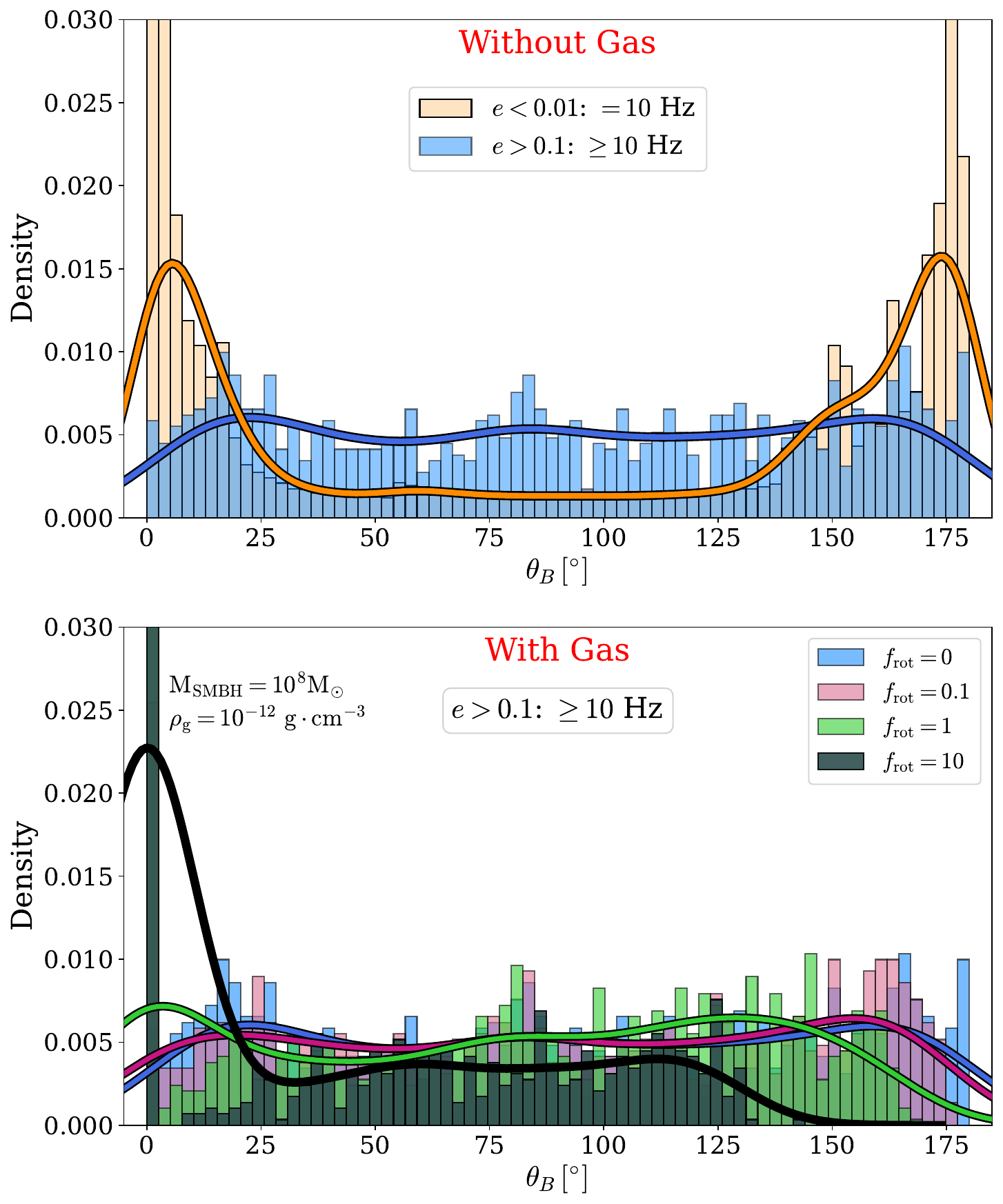}
    \caption{Distribution of binary inclinations $\theta_B$ for different merger populations.
    \textit{Top}: results without gas. Quasi-circular mergers ($e<0.01$ at 10 Hz, yellow) show sharp peaks at $0^\circ$ and $180^\circ$, while eccentric mergers ($e>0.1$, blue) are more isotropic with a flatter distribution. Solid curves indicate the smoothed probability density estimates.
    \textit{Bottom}: results including gas alignment for a $10^8 M_\odot$ SMBH and disk density $\rho_{\rm g}=10^{-12} \, \rm g \, cm^{-3}$ (representative of the \citetalias{Sirko03} model), with alignment efficiencies $f_{\rm rot}$ = 0 (blue), 0.1 (pink), 1 (green), and 10 (black). Increasing $f_{\rm rot}$ reduces the fraction of retrograde binaries,
    and at high efficiency ($f_{\rm rot} = 10$) the distribution departs significantly from the gas-free case. The evolution of $\theta_B$ is quantified using the expression given in Eq.~\eqref{eq:efold}.
    Histograms are based on PN N-body simulations of equal-mass $15 \, \rm M_\odot$ BHs with $a_0 = 1$ AU, where the single approaches the binary at an initial inclination of $0.1^\circ$.}
    \label{fig:idist}
\end{figure*}

\subsection{Mapping Alignment Based on Disk Location} \label{sec:map}

Fig.~\ref{fig:map} presents contour maps of the alignment timescale across the \citetalias{Sirko03} disk, shown as a function of SMBH mass and distance from the SMBH in units of $R_{ s}$. The timescales are computed for an equal-mass binary with individual components of $M_{\rm BH} = 15 \, \rm M_{\odot}$ and an initial semi-major axis of $a_0=1$ AU. The top row corresponds to binaries with an initial inclination $\theta_B = 10^{\circ}$, while the bottom row shows the case of $\theta_B = 110^{\circ}$. 

The alignment timescales are shown for three representative values of the eccentricity at 10 Hz ($e_{\rm 10Hz} = 0.01, 0.1, 0.9$). Note that alignment time is evaluated at dynamical formation, when binaries are still extremely eccentric, but $e_{\rm 10Hz}$ provides a useful way of classifying the populations observable in the LVK band. The shaded black regions indicate where $a_0$ exceeds half the Hill radius. Such cases are excluded, as we only consider binaries that remain well embedded within their Hill sphere.
To directly assess whether alignment can occur before merging, we compare $t_{\rm align}$ to the GW inspiral timescale, $t_{\rm gw}$. The solid black contour lines denote $t_{\rm align}/t_{\rm gw} = 1$. 

Different line-styles correspond to alignment efficiencies: $f_{\rm rot} = 0.1$ (solid; fiducial), 1 (dashed), and 10 (dot-dashed). In addition, dotted contours highlight the region where $t_{\rm align}(f_{\rm rot}=0.1) < t_{\rm gw}$, corresponding to cases where the binary fully aligns with $\hat{J}_{\rm AGN}$ before merger. As expected, increasing $f_{\rm rot}$ enlarges this region of parameter space, since alignment becomes increasingly more efficient. 
A few features stand out from these maps. First, as evident from Eqs.~\eqref{eq:ratioad}~\&~\eqref{eq:ratiotl}, alignment scales linearly with the disk density. Consequently, alignment becomes less effective at larger radii where the disk is puffier and less dense, leading to longer timescales. Second, nearly circular binaries ($e_{\rm 10Hz} \lesssim 0.01$) realign efficiently across most of the parameter space, for both low and high inclinations. This suggests that quasi-circular mergers will generally be observed with spins aligned to the disk, regardless of their initial orientation. 

In contrast, as $e_{\rm 10Hz}$ increases, the competition between $t_{\rm align}$ and $t_{\rm gw}$ becomes much sharper: the region where alignment dominates shrinks substantially. For highly eccentric sources, particularly at $\theta_B = 110^{\circ}$, the GW inspiral timescale is shorter than the alignment timescale. This implies that such binaries are expected to merge with significant misalignment even in the presence of gas accretion. This behavior is consistent with the analytic scaling, which shows that when $\tau_{\rm align}$ and $t_{\rm gw}$ are of similar order, the $(1-e_0^2)^{-3}$ dependence drives $t_{\rm gw}$ down by orders of magnitude at high eccentricity.
Finally, it is important to emphasize the role of disk parameters in shaping these results. The maps in Fig.~\ref{fig:map} are generated assuming the fiducial \citetalias{Sirko03} model parameters, with viscosity $\alpha = 0.01$ and Eddington fraction $l_{\rm E}=0.5$. Modifying these parameters directly affects the density and sound speed of the disk, and therefore the efficiency of alignment. Increasing $\alpha$ in the \citetalias{Sirko03} framework lowers the density and temperature in the inner regions, thereby suppressing accretion and reducing alignment efficiency. On the contrary, lowering $l_{\rm E}$ reduces the accretion rate, which decreases the sound speed but raises the density, leading to stronger alignment torques. \footnote{These effects occur in the \citetalias{Sirko03} model because enforcing marginal stability forces the disk to adjust its surface density and temperature to maintain $Q\sim 1$.}
This underscores the fact that disk properties and initial conditions are critical when assessing whether alignment or GW inspiral dominates, and that obtaining a universal conclusion is non-trivial.

\begin{figure}
    \centering
    \includegraphics[width=\linewidth]{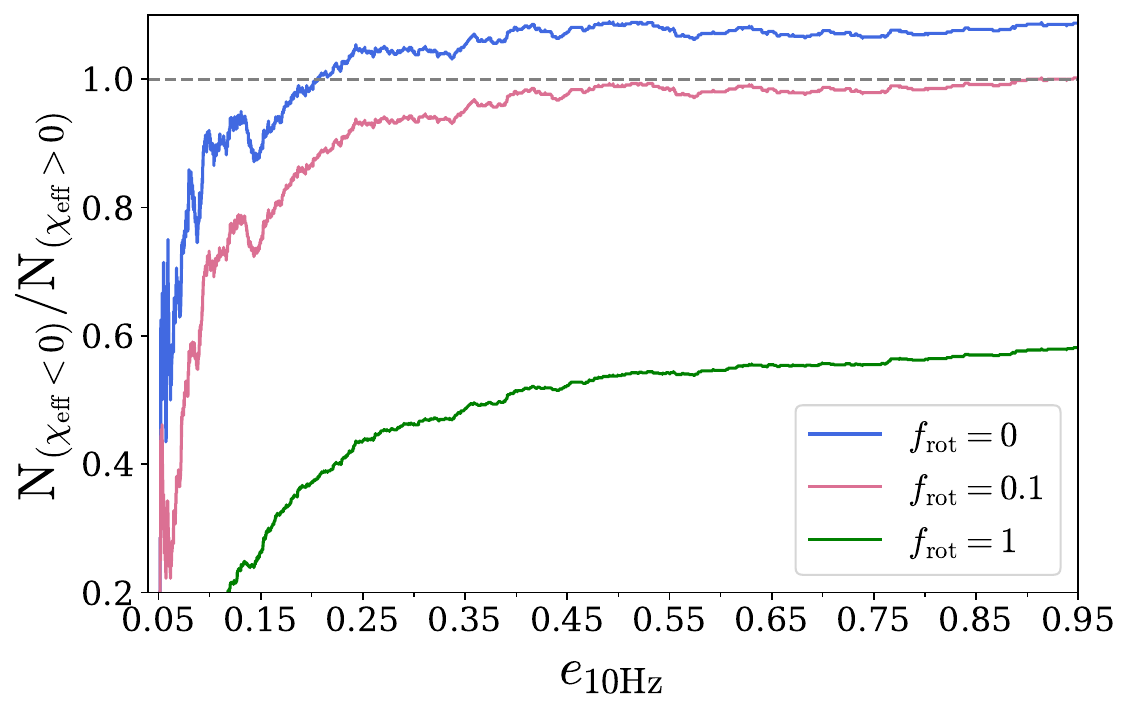}
    \caption{Ratio of negative to positive effective spin orientations, 
    $N_{(\chi_{\rm eff}<0)}/N_{(\chi_{\rm eff}>0)}$, as a function of eccentricity at 10 Hz ($e_{\rm 10Hz}$). 
    Results are shown for different alignment efficiencies: 
    $f_{\rm rot}=0$ (no alignment, blue), $f_{\rm rot}=0.1$ (pink), and $f_{\rm rot}=1$ (green). 
    For small eccentricities and high $f_{\rm rot}$, alignment is efficient and the ratio remains well below unity, 
    reflecting a distribution weighted toward positive values of $\chi_{\rm eff}$ in AGN disks. 
    For $f_{\rm rot}=0.1$ and at larger $e_{\rm 10Hz}$, the ratio grows and saturates near unity, 
    indicating that alignment becomes less efficient than the GW inspiral. 
    Without the effect of alignment, the excess of retrograde binaries reflects the outcome of three-body dynamics, while larger encounter angles ($\varphi$) suppress their production.}
    \label{fig:cumdist}
\end{figure}
\subsection{Spin-Orbit Tilt Distribution and Evolution } \label{sec:spinev}

The upper panel of Fig.\ref{fig:idist} shows the distribution of orbital orientations $\theta_B$ obtained from 2.5PN three-body scattering experiments, as explained in Sec~\ref{sec:scatterings}. We start by considering results without the effect of gas alignment. 
As discussed in Sec.\ref{sec:scatterings}, this work focuses on out-of-plane interactions, where the incoming single approaches the binary in the disk mid-plane at a small inclination angle ($\varphi = 0.1^\circ$ in this case). The simulations are performed for a BBH with an initial semi-major axis of 1 AU. For each experiment, we keep $a_0$ and $\varphi$ fixed while varying the binary orbital phase (i.e. the position of the binary along its orbit at the time of the encounter). 
The angle $\theta_B$ is measured after the binary–single interaction has taken place (middle panel of Fig.\ref{fig:sketch}).

The normalized distribution in the upper panel is presented for two distinct merger populations. Quasi-circular sources, defined as binaries with eccentricity $e<0.01$ at a GW peak frequency of 10 Hz, are shown in yellow, while eccentric mergers with $e>0.1$ at 10 Hz are shown in blue. The solid curves show smoothed probability density estimates, computed using a Gaussian kernel to trace the overall distribution trend of the histogram for each case. 
The quasi-circular population displays sharp peaks at $\theta_B \simeq 0^\circ$ and $180^\circ$, while the eccentric mergers exhibit a much flatter, nearly isotropic distribution. The origin of this difference, as pointed out in \citet{Samsing22}, lies in the initial angular momentum 
content of the two populations: for a fixed semi-major axis, eccentric binaries have smaller orbital angular momentum vectors making them easier 
to tilt during the encounter, while quasi-circular mergers, with their larger angular momentum, are harder to re-orient and thus tend to remain co-planar.

As shown in Fig.~\ref{fig:map}, once the effect of accretion is introduced, quasi-circular binaries will typically re-align across most of the parameter space, independently of AGN properties or initial binary orientation. We therefore expect the quasi-circular population to remain preferentially aligned with $\hat{J}_{\rm AGN}$ by the time they merge. This scenario, where quasi-circular mergers produce sharp peaks at prograde and retrograde orientations, 
may be applicable in quiet galactic nuclei with a nuclear stellar disk, where alignment torques are weaker. Nevertheless, even in such environments, previous work suggests that spin orientations can still efficiently re-align through interactions with massive stars \citep{2025ApJ...983L...9K}.

In contrast, for a significant fraction of the parameter space, eccentric binaries are expected to merge before the alignment process is complete. 
The lower panel of Fig.~\ref{fig:idist} shows the distribution of $\theta_B$ for the eccentric population taking gaseous effects into consideration,  
choosing representative values from the \citetalias{Sirko03} model for a $\rm 10^8 \, M_\odot$ SMBH and a disk density of $\rho_{\rm g} = 10^{-12}\,\rm g\,cm^{-3}$.
We quantify the alignment by tracking how much the binary’s inclination angle $\theta_B$ evolves during the inspiral phase. Specifically, 
we integrate Eq.~\eqref{eq:diff} up to the merger time, setting the upper limit of integration to the gravitational wave inspiral timescale $t_{\rm gw}$. This yields an expression for the inclination at merger:
\begin{equation}\label{eq:efold}
    \theta_{B, \rm merge} (t) = \cos^{-1} \left(\cos \theta_{B,0}+\frac{t}{\tau_{\rm align}}  \right),
\end{equation}
where $\theta_{B,0}$ is the post-scattering orientation and we evaluate the final angle at $t = t_{\rm gw}$. When $t_{\rm gw} > \tau_{\rm align}$, we consider $\theta_B$ to have been fully aligned to 0, regardless of the initial inclination.
Eq.~\eqref{eq:efold} therefore provides a direct way to connect the initial misalignment at formation to the inclination at merger, given a prescribed alignment efficiency.

It is important to stress the assumptions implicit in this treatment. 
First, when evaluating Eq.~\eqref{eq:efold} we adopt the binary separation $a_0$ at the moment of formation. 
This ignores the fact that gas can drive rapid orbital shrinkage \citep[e.g.][]{Baruteau:2010bk,Li22c,Rowan23b,Rowan2025_triples}, which would reduce the merger time and shrink the window for alignment.
Moreover, $e_0$ can also be modified by gas: if gas excites eccentricity in the binary, its merger time decreases even more steeply, further limiting alignment. On the contrary, if gas damps eccentricity, the inspiral timescale slows down, allowing alignment to proceed more efficiently. Simulations suggest that this behavior is determined by the orientation, where retrograde binaries exhibited eccentricity growth and vice versa for prograde binaries \citep[e.g][]{Rowan23b,2024ApJ...970..107C,Dittman2025_unequal_m}.
A more realistic calculation would couple the time evolution of $a_0$ and $e_0$ to the tilt evolution, 
solving the full system of differential equations rather than applying a single timescale, but at present this remains intractable.
Second, as also observed in simulations \citep[e.g.][]{Dittmann24}
Eq.~\eqref{eq:efold}, like Eq.~\eqref{eq:talign}, assumes that retrograde binaries 
always evolve toward prograde orientations rather than becoming more retrograde, even for near-$180^\circ$ inclinations. 
The physical motivation for this choice is discussed further in Sec.~\ref{sec:retro}.

The distributions in the lower panel of Fig.~\ref{fig:idist} are shown for different values of alignment efficiency $f_{\rm rot}$: 
0 (blue, no alignment), 0.1 (pink), 1 (green), and 10 (black).
Similar to the upper panel, the solid curves show the smoothed probability density estimates. As expected, with increasing value of $f_{\rm rot}$, the number of retrograde binaries rapidly decreases, and for the highest value of $f_{\rm rot}$, the distribution peaks at 0, and the number of $\theta_B >90^{\circ}$ significantly decreases.
We find that, for a selected range of disk densities, in order for the distribution to differ substantially from the gas-free case, the alignment efficiency must be $\sim 2$ orders of magnitude higher than our fiducial value $f_{\rm rot}=0.1$, which is calibrated to recent hydrodynamical simulations of inclined binaries \citep{Dittmann24}.
Finally, we note that the $f_{\rm rot}=10$ case may not be physically achievable in AGN disks: the corresponding accretion rates are already highly super-Eddington in our setup, and increasing the effective torque efficiency by two orders of magnitude is difficult to justify. Addressing whether such high values of $f_{\rm rot}$ can be realized will require dedicated  simulations. 

We also present results spanning the full range of eccentricities, focusing in particular on how the 
distribution of the effective spin parameter $\chi_{\rm eff}$, evolves depending on binary eccentricity at 10 Hz 
($e_{\rm 10Hz}$) and on alignment efficiency $f_{\rm rot}$. We therefore examine whether the distribution of 
$\chi_{\rm eff}$ is preferentially weighted toward positive (aligned) or negative (anti-aligned) values. 
As already discussed in Sec.~\ref{sec:tilt_methods}, although the orbital inclination $\theta_B$ is not directly 
observable, if the gas in the disk aligns the individual BH spins with the disk angular momentum 
\citep[e.g.][]{Tagawa20}, then the effective spin parameter becomes directly related to the binary tilt through 
$\chi_{\rm eff} \propto \cos\theta_B$ \citep[e.g.][]{Samsing22}. We therefore classify binaries with 
$\theta_B \leq 90^{\circ}$ as ``positive'' $\chi_{\rm eff}$ systems ($\chi_{\rm eff} > 0$), and those with 
$\theta_B > 90^{\circ}$ as ``negative'' $\chi_{\rm eff}$ systems ($\chi_{\rm eff} < 0$). 

Figure~\ref{fig:cumdist} shows the ratio between the number of negative and positive $\chi_{\rm eff}$ binaries, 
$N_{\chi_{\rm eff}<0}/N_{\chi_{\rm eff}>0}$, as a function of eccentricity at 10 Hz spanning between 0.05 and 0.95, for three values of alignment efficiency: 
$f_{\rm rot}=0$ (blue, no alignment), $f_{\rm rot}=0.1$ (pink), and $f_{\rm rot}=1$ (green). 
As expected from the $(1-e^2)^{-3}$ scaling in Eqs.~\ref{eq:ratioad}--\ref{eq:ratiotl}, which sets the ratio 
$\tau_{\rm align}/t_{\rm gw}$, alignment is most effective for low eccentricities and large $f_{\rm rot}$, 
yielding $N_{\chi_{\rm eff}<0}/N_{\chi_{\rm eff}>0} \ll 1$.  

For our fiducial value of $f_{\rm rot} = 0.1$, at larger $e_{\rm 10Hz}$ the ratio increases significantly and eventually plateaus very close to unity. 
In the absence of alignment, the ratio exceeding unity merely reproduces the initial distribution generated by the three-body encounter,
where retrograde binaries are more common. This outcome reflects the dynamical nature of the encounters rather than any spin-related effect.
Increasing the inclination of the incoming single, $\varphi$, systematically lowers the retrograde fraction, as larger approach angles diminish the efficiency of plane-flipping interactions and thus reduce the number of retrograde outcomes.

Finally, for higher alignment efficiency ($f_{\rm rot}=1$), the distribution of $\chi_{\rm eff}$ values becomes strongly 
positive, recovering the one typically expected for AGN-assisted mergers where gas torques dominate \citep[e.g.][]{Tagawa20,Mcfacts1,2025arXiv250608801F}. 
Taken together, these results indicate that the interplay between binary eccentricity, disk density, and 
alignment efficiency $f_{\rm rot}$ controls whether the $\chi_{\rm eff}$ distribution in AGN disks is biased toward positive as generally expected or if its values remain closer to isotropic. 
The systems that merge before full alignment, are in addition expected to exhibit enhanced in-plane spin components $\chi_P$ \citep{2025arXiv250608801F}. In these cases, the effective spin can remain small even while the total spin magnitude is large, thus providing a complementary diagnostic of possible distinctive features of the AGN channel. 

\subsection{Upper Limit for the Alignment Angle}\label{sec:uplim}

\begin{figure}
    \centering
    \includegraphics[width=\linewidth]{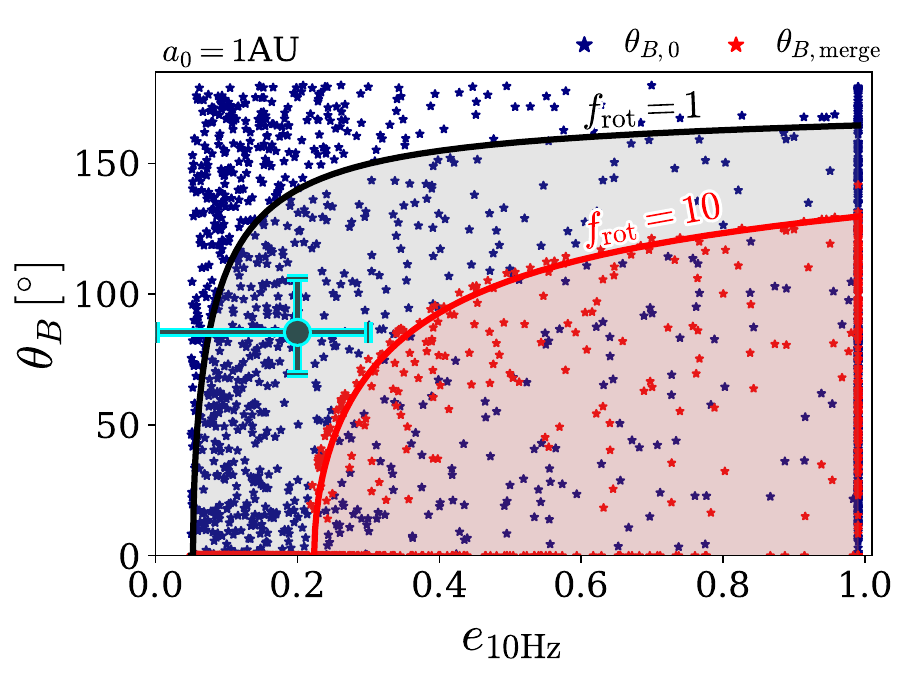}
    \caption{Initial (blue stars) and final (red stars) inclination distributions $\theta_B$ as a function of $e_{\rm 10Hz}$, with the final points shown for $f_{\rm rot}=10$. The red curve represents the maximum final inclination available for a binary starting at $\theta_{B,0}=180^\circ$, and the shaded region illustrates the available combinations of $\theta_{B,\, {\rm merge}} - e_{\rm 10Hz}$ at $f_{\rm rot}=10$.
    We also show the limiting curve for $f_{\rm rot}=$ 1 (black line), together with the associated black shaded region of available final combinations.
    For the higher alignment efficiency case, binaries with $e_{\rm 10Hz} \lesssim 0.2$ are fully realigned, while higher eccentricities lead to progressively larger residual tilts. The most eccentric mergers, with GW peak frequencies already above 10 Hz at formation, retain significant retrograde orientations due to their extremely short merger timescales.
    The green point shows GW190521. The inclination is inferred from the reported $\chi_{\rm eff}$,  while the eccentricity estimate is taken from \citet{2022NatAs...6..344G} }.
    \label{fig:appendix_icrit}
\end{figure}

In this section we further explore how the efficiency of 
gas-driven alignment depends on the eccentricity at 10 Hz. 
Fig.~\ref{fig:appendix_icrit} displays the initial distribution of binary orientations, $\theta_{B,0}$ (blue stars),
and the corresponding final orientations at merger, $\theta_{B,\rm merge}$ (red stars), shown for the $f_{\rm rot}=10$ case.
The red curve indicates the maximum final alignment angle for a binary starting at $\theta_{B,0}=180^\circ$
(from Eq.~\ref{eq:efold}), as a function of $e_{\rm 10Hz}$, and the shaded red region illustrates the set of allowed
$\theta_{B,{\rm merge}}-e_{\rm 10Hz}$ combinations for this torque efficiency.
To illustrate how this behavior changes at lower torque efficiency, we also show the corresponding limiting curve
and shaded region for $f_{\rm rot}=1$ in black.

Fig.~\ref{fig:appendix_icrit} shows that for the highest value of $f_{\rm rot}$ binaries with $e_{\rm 10Hz} \lesssim 0.2$ are efficiently aligned regardless of their initial inclination, but for lower values of $f_{\rm rot}$ nearly the entire space remains possible. For both torque efficiencies the maximum post-alignment angle increases with eccentricity, further indicating that eccentric binaries are more likely to retain a spin-orbit misalignment at merger. In particular, the binaries with peak frequency above 10 Hz at formation also inspiral so rapidly that they can retain significant retrograde orientations despite strong alignment torques. The efficiency of this process, though, always depends sensitively on the local disk density: in denser regions alignment can occur much faster as $\tau_{\rm align} \propto  \rho_{\rm g}$. 

The figure also includes GW190521, shown as a large green data-point with its associated uncertainties. 
The inclination is inferred from the reported $\chi_{\rm eff}$ in \citet{GW19a} assuming $\chi_{\rm eff}\propto\cos\theta_B$, while the eccentricity estimate is taken from \citet{2022NatAs...6..344G}. 
Under these assumptions, GW190521 lies in a region of parameter space consistent with $f_{\rm rot}\sim 1$. 
More generally, the framework presented here can be applied to individual merger events by mapping constrained values of eccentricity and effective spin onto the predicted regions of the $e_{\rm 10Hz}-\chi_{\rm eff}$ plane. 
As illustrated in Fig.~\ref{fig:appendix_icrit}, even single-event measurements with large uncertainties can be used to assess whether a merger is consistent with rapid inspiral and incomplete alignment, or with more efficient alignment in the circular regime. 
We can therefore identify which values of the torque efficiency parameter $f_{\rm rot}$ are consistent with a given event. Since $f_{\rm rot}$ depends on local disk conditions, assuming a disk origin, we can also probe the local disk density at the merger site. 
Such applications will become increasingly powerful as eccentric waveform models improve and detector sensitivity increases.

\subsection{Retrograde orbiters and warp radius}\label{sec:retro}

The alignment of isolated binary-disk systems (as opposed to disk-embedded binaries treated here) is typically governed by an analog to the Bardeen-Petterson effect whereby differentially precessing annuli, whose angular momentum vectors are tilted relative to that of the binary, are viscously damped into the plane of the binary orbit \citep{BP75, Nixon11}.\footnote{Provided the disk is sufficiently viscous.} 
This is effective out to the warp radius,
\begin{equation}\label{eq:warp}
    R_{\rm warp} = \frac{2 G J}{c_s^2 \nu_2} 
    = \frac{2 G m_{\rm bin}}{c_s^2} 
    \frac{\sqrt{G m_{\rm bin} a_0}}{\nu_2},
\end{equation}
where the vertical shear viscosity $\nu_2$ is set equal to the standard $\alpha$-parametrized azimuthal viscosity, $\nu_2 = \alpha\,c_s\,H$, computed from the local disk properties.
This is a simple model for the structure of the accretion flow that does not account for local deviations from the ambient disk \citep[e.g.,][]{2022ApJ...928L..19L}.
At the warp radius, the viscous time associated to $\nu_2$ becomes equal to the local precession time \citep{Pringle:1992, ScheuerFeiler:1996, King05}.
For disk-embedded binaries, if the accretion flow were sufficiently independent from the large scale shear of the AGN-disk, the associated timescale could alternatively be applied to estimate $\tau_{\rm align}$.

\begin{figure}
    \centering
    \includegraphics[width=\linewidth]{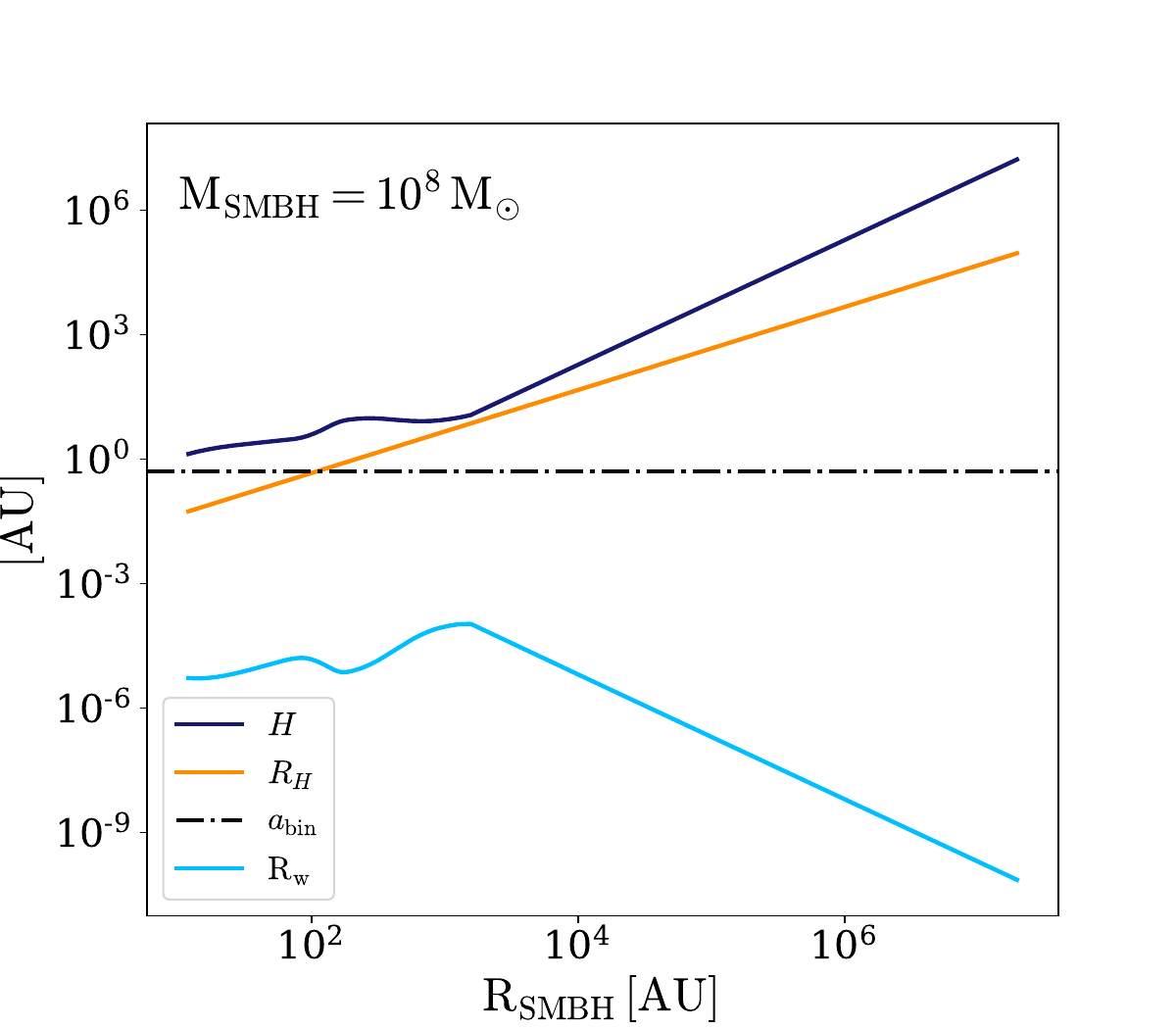}
    \caption{Characteristic length scales (in AU) for a
    $10^8 \, M_\odot$ SMBH and a binary with equal mass
    $15 \, M_\odot$ components and $a_0 = 0.5$AU. Shown are the disk scale height 
    $H$ (blue), the Hill radius $R_H$ (orange), and the warp radius 
    $R_{\rm warp}$ (light blue) computed from Eq.~\eqref{eq:warp}. 
    The horizontal dash-dotted line marks the fiducial binary semi-major 
    axis, $a_0 = 0.5 \, \rm AU$. 
    The warp radius is orders of magnitude smaller than $a_0$, so binary alignment is tidal and standard retrograde stability \citep[e.g.,][]{King05, Nixon11} are not present.
    }
    \label{fig:radii}
\end{figure}

This process also forms the basis for a (temporarily) stable retrograde phase in which the disk interior to $R_{\rm warp}$ counter-aligns with $\vec{J}$ if $\cos{\theta_b} < - J_d(R_{\rm warp}) / (2 J)$, where $J_d(R_{\rm warp})$ is the disk angular momentum out to $R_{\rm warp}$.
To explore the possibility of a stable retrograde phase, we plot in Fig.~\ref{fig:radii} the warp radius (blue) as a function of the binary distance from the central SMBH, alongside the typical disk scale height $H$ (purple) and Hill radius (orange) for fiducial binary and disk parameters.
We assume that the vertical viscosity is comparable to the azimuthal viscosity, $\sim \alpha c_s H$, and that these quantities correspond to their local values in the SG disk solution.
We find that while the warp radius falls within the binary's Hill sphere and the disk's scale height, it is many orders of magnitude smaller than the binary separation itself (i.e., $0.5 $AU).
As a result, the embedded binary cannot locally align material into its orbital plane, the alignment process is tidal as opposed to viscous, and initially nearly-retrograde inclined binaries would not be expected to counter-align with $\hat J_{\rm AGN}$.
This agrees with the previous assumptions for the form of $\dot J_z$ and that all binaries evolve toward alignment with the disk angular momentum.
Of note, the simulations in \citet{Dittmann24} find that mini-disks around the individual BHs of a misaligned binary are aligned with $\hat J_{\rm AGN}$, not $\hat J$.\footnote{Where the spins of the black holes themselves are not treated.} 
We posit that this is because even the characteristic mini-disk size ($\sim 0.3 \, a_0$) far exceeds the warp radius such that the gas does not have sufficient time to align with the orbit as it is captured around an individual black hole.
This bastions the claim that, while the orbit can be mis-aligned to the disk, the spins ought generally to align with $\hat{J}_{\rm AGN}$ (but see also discussion of caveats below).

\section{Caveats} \label{sec:caveats}

While our analysis provides a first step toward understanding the alignment of eccentric mergers in an AGN disk, it remains simplified on several aspects. Here we outline the main limitations and highlight where more sophisticated tools (e.g. hydrodynamical simulations) and more detailed modeling will be essential. It is important to stress that these assumptions are not simply modeling choices for simplicity: rather, they reflect outstanding uncertainties in the astrophysical processes themselves, 
where theory and simulations have either not addressed the problem yet or have not converged on a definitive answer.

\begin{itemize}
    \item \textbf{Gas effects and time-dependent evolution}: In this work we do not include the effect of binary-disk interactions on eccentricity and semi-major axis, treating gravitational radiation as the sole driver of orbital decay and eccentricity damping. In principle, gas can modify the inspiral by shrinking $a_0$ and damping or exciting $e_0$, depending on whether the orbit is prograde or retrograde \citep[e.g.][]{2024ApJ...970..107C,2024ApJ...974..216O}. A more realistic treatment would require solving the coupled differential equations for orbital evolution under the combined influence of $\dot{a}_{\rm gw}+\dot{a}_{\rm gas}$ and $\dot{e}_{\rm gw}+\dot{e}_{\rm gas}$. 
    We comment, though, that we do not expect our general prediction that eccentric binaries are more likely to merge mis-aligned (while circular ones will merger aligned) to change substantially because of the scalings with $a$ and $e$ in Eqs.~\eqref{eq:scaling_ad}~\&~\eqref{eq:scaling_tl}. For eccentricity in particular, this is because gas-driven eccentricity pumping (unless coupled to orbit expansion) will decrease the available alignment time, while eccentricity damping will simply create more circular sources.
    Furthermore, hydrodynamical simulations can provide an additional tool to explore this coupled evolution and to determine whether our prescription under- or overestimates the effective efficiency of alignment. Verifying this coupling is especially important for eccentric binaries, where rapid orbital shrinkage significantly limits the time window over which gas can act. At present, however, the correct form and expected results of the coupled $\dot{a},\dot{e}$ evolution in AGN disks is still under major study and uncertain, which is why we do not include it in our analysis.
    \item Accretion onto the individual BBH components: 
    A key assumption of our analysis is that only a limited fraction of the ambient disk gas flows into the binary, such that gas--binary coupling primarily manifests through gravitational torques rather than substantial BBH mass growth. 
    Recent hydrodynamical simulations indicate that accretion onto embedded binaries is highly time-variable, with orbital evolution determined by complex gas dynamics and gravitational torques, rather than by a simple, steady accretion flow \citep[e.g.,][]{Li22a,Li22c}. 
    These studies further indicate that the gas distribution near the binary is strongly non-axisymmetric \citep{Rowan23b} and therefore only a fraction of the surrounding gas might become bound to the individual black holes. 
    If, however, circum-binary flows were found to be highly efficient at feeding the binary, the resulting growth of the black hole masses would shorten the gravitational-wave merger timescale (Eqs.~\eqref{eq:ratioad}--\eqref{eq:ratiotl}). 
    In this regime, binaries could merge before alignment is completed, even for circular systems. 
    While this does not affect our conclusion that eccentric binaries typically merge before alignment due to their short gravitational-wave timescales, it could reduce the fraction of aligned circular mergers and weaken the predicted correlation between eccentricity and $\chi_{\rm eff}$ in selected parts of the disk.

    \item \textbf{Super-Eddington accretion:} 
    Our prescription assumes accretion rates that are often highly super-Eddington, raising the question of whether such flows can be sustained in AGN disks. If that is not the case, alignment efficiencies would be significantly reduced, as $\dot{M}$ would occur at a much slower rate, unless $f_{\rm rot}$ was raised commensurately. 
    There remains substantial uncertainty as feedback from radiation pressure and jet-driven outflows could strongly regulate or even suppress such accretion \citep{Pan2021,Tagawa2022_accretion}. 
    It is therefore important to understand whether AGN disks can sustain these regimes, since the physical justification of large alignment efficiencies (e.g. $f_{\rm rot} \gtrsim 1$) would also depend on the availability of super-Eddington accretion. 
    If the actual accretion rate is much lower than assumed in our models, this would imply that many of our circular binaries would also not have time to align, and shift the anticipated $\chi_{\rm eff}$ distribution of the AGN channel more towards zero. 
    Whether sustained super-Eddington accretion is possible in AGN disks is still an unresolved problem, and thus our treatment still represents one plausible scenario.
    \item \textbf{Scattering in and out of the disk:} In our alignment prescription we assume that binaries remain confined to the disk after formation. In reality, repeated encounters can scatter black holes out of the disk plane, temporarily removing them from the dense gaseous environment where alignment operates most efficiently. Moreover, as argued by \citet{2025arXiv250608801F}, the deep SMBH potential well typically retains BHs even after they are scattered out of the disk, causing them to remain bound and re-cross the disk on orbital timescales. The net impact on the inclination distribution therefore depends on how often and on which timescales systems leave and re-cross the disk.
    \item \textbf{Inclination distributions including gas drag:} our results are based on pure N-body simulations with the inclusion of the SMBH in the equations of motion but not of the effect of the gaseous drag force. For the initial inclination distribution at merger, gas drag can potentially alter these outcomes by e.g. damping eccentricities in the case of prograde orbits. This would reduce the fraction of binaries entering the LVK band with very high $e_{\rm 10Hz}$ and can potentially broaden the population of quasi-circular mergers. Incorporating the effect of gas into the scattering experiments is important to provide further insights on the actual distribution of inclinations after the interaction, as the extent to which gas modifies binary inclination distributions is still largely unexplored. 
    \item \textbf{Turbulence}: throughout the paper we neglect the effect of turbulence. Turbulence in AGN disks can yield stochasticity in the accretion process and randomize angular momentum transport \citep{Chen2023_chaotic_acc, 2024arXiv240905614M}, effectively reducing the efficacy of alignment \citep{2025arXiv250602173T}. In this case, even binaries embedded in the densest regions of the disk might experience weaker alignment than predicted. 
    Turbulent accretion and supersonic motion relative to the disk can also misalign or flip mini-disks, potentially decoupling the individual black-hole spins from $\mathbf{J}_{\rm AGN}$ \citep{2022ApJ...928L...1L,Chen2023_chaotic_acc}.
    The details on strength effect of turbulence in AGN disks and its role in shaping the evolution of BBHs remain an open problem in astrophysics. 

    \item \textbf{Choice of disk model}: 
    The alignment timescale has a strong dependence on the local disk properties and therefore on the adopted AGN disk model.
    In this work, we focus on the \citetalias{Sirko03} model to provide a representative example, particularly because it is characterized by relatively high densities in the inner regions, where alignment is most efficient. 
    Selecting a different model such as the \citet{Thompson05}, which predicts lower inner densities compared to \citetalias{Sirko03},
    would yield longer characteristic alignment timescales and more pronounced spin-orbit misalignment for eccentric sources.
    \item \textbf{Repeated interactions and hierarchical mergers:} In our $N$-body simulations, the integration is terminated once the binary enters the inspiral phase. The majority of these mergers are three-body mergers, in which the third companion remains bound to the system at the time of merger assembly. This raises the possibility of further dynamical interactions: while the binary is inspiraling and potentially re-aligning through accretion, the bound single may return and perturb the system again. Even if the single is temporarily ejected beyond the Hill sphere during the initial encounter, the presence of the AGN disk favours its re-formation \citep[e.g.][]{Rowan2025_triples}, enabling subsequent close passages. Most importantly, the single is not only capable of re-entering the system and scattering the binary during its inspiral, but it may also merge with it, leading to a hierarchical merger \citep{Samsing18a,2019MNRAS.482...30S}. Recent studies have even suggested that distinct gravitational wave events could be connected through their proximity in sky localization, raising the possibility that one merger directly followed another within the same AGN environment \citep{Veske2022,2025PhRvD.111j3016L}.
    Such repeated scatterings during the inspiral phase could alter the orbital configuration of the merging binary, affecting the realignment process. Although our present simulations do not follow this process beyond the initial inspiral, exploring the combined role of repeated encounters, gas-assisted recapture, and the possibility of hierarchical mergers will examined future work.

\end{itemize}

\section{Conclusions} \label{sec:conclusion}

In this section, we summarize the main findings of this work. Our study combined 3D PN N-body simulations of binary–single scatterings with an analytical prescription for gas-driven alignment in AGN disks, in order to map how spin–orbit tilts evolve from formation to merger. A key focus is the competition between alignment and gravitational wave inspiral, which determines whether binaries are able to re-align with the disk before coalescence, or if they instead merge with significant misalignments. We also investigate how these dynamics shape the distribution of binary inclinations, and by extension the effective spin parameter $\chi_{\rm eff}$. 

Finally, we discuss the broader implications of our results for gravitational wave observations, including the interpretation of individual events such as GW190521 and GW231123, and the potential signatures of the AGN channel in the overall LVK population.

\begin{itemize}

    \item \textbf{The alignment process:} We develop an analytical prescription for the alignment of black hole binaries embedded in AGN disks, focusing on how gas accretion modifies the in- and out-of-plane components of the binary angular momentum, testing different accretion efficiencies through the parameter $f_{\rm rot}$. We explore the different regimes inside an AGN disk (accretion-dominated and tidally limited), which depend on whether the relevant accretion scale is set by the Bondi or Hill radius. This framework allows us to systematically connect the binary’s orbital properties (semi-major axis, eccentricity, inclination) to the surrounding disk conditions (density, sound speed), and to track how efficiently gas-driven torques can realign the system before it merges. We find that alignment is most effective in dense, inner disk regions, and it is sensitive to the binary’s orbital eccentricity and the local accretion efficiency.

    \item \textbf{Comparison of alignment and GW inspiral timescales:} In Sec.~\ref{sec:map} we explore the interplay between alignment and GW-driven inspiral across different disk locations and properties, adopting the \citetalias{Sirko03} model and  sampling different binary eccentricities at 10 Hz and initial inclinations. Alignment timescales scale linearly with both disk density and $f_{\rm rot}$, whereas the GW merger timescale depends steeply on eccentricity and semi-major axis. As a result, in regions where the disk density is insufficient for alignment to dominate, compact and highly eccentric binaries inspiral so rapidly that they preserve their spin–orbit tilt. In contrast, nearly circular binaries experience efficient realignment across most of parameter space. This comparison establishes a framework for identifying the regimes in which binaries are expected to merge with substantial spin–orbit tilts versus those in which they are re-aligned with the AGN disk.
    
    \item \textbf{Distribution of binary inclinations with and without alignment:} Using PN $N$-body simulations, we first map the inclination distributions of binaries assembled through three-body scatterings, and then evolve them under the influence of gas alignment. In the absence of alignment, we find two distinct populations: quasi-circular mergers, which display peaks around $0^\circ$ and $180^\circ$, and eccentric mergers, which display a flatter, nearly isotropic distribution due to their smaller angular momentum vectors being more easily tilted (as already pointed out in \citealt{Samsing22}). Once gas effects are included, quasi-circular binaries re-align efficiently across most of the parameter space, regardless of their initial inclination, and are therefore expected to merge aligned with the disk. By contrast, eccentric binaries can often merge before alignment is complete if the density is low ($\rho_{\rm g} \lesssim 10^{-12} , \rm g , cm^{-3}$) or the efficiency modest ($f_{\rm rot} \lesssim 1$). To follow this evolution we solve Eq.~\eqref{eq:efold} up to the merger time, measuring how $\theta_B$ decays as a function of $f_{\rm rot}$. Increasing $f_{\rm rot}$ progressively suppresses retrograde systems, and at $f_{\rm rot}=10$ the distribution peaks near $0^\circ$ with retrograde binaries nearly absent. At the fiducial efficiency $f_{\rm rot}=0.1$, however, the distributions remain close to the gas-free case. 
    
    \item \textbf{Correlation between eccentricity and effective spin in the AGN channel:}
    Although $\theta_B$ is not directly measurable, if gas aligns the individual BH spins with the disk angular momentum \citep[e.g.][]{Tagawa20,Avi2022}, then $\chi_{\rm eff}$ follows $\cos\theta_B$ \citep[e.g.][]{Samsing22}, allowing us to link orbital tilt to the effective spin. We find that the distribution of $\chi_{\rm eff}$ values depends strongly on eccentricity and alignment efficiency. Quasi-circular binaries ($e_{\rm 10Hz}\lesssim 0.01$) consistently re-align and yield positive $\chi_{\rm eff}$. Eccentric binaries, on the other hand, can retain isotropy unless alignment is sufficiently effective: at our fiducial value $f_{\rm rot}=0.1$, the distribution remains nearly flat, but for higher efficiencies ($f_{\rm rot}\gtrsim 1$) or denser disks, it recovers a positive bias. This provides a natural way to map a correlation between binary eccentricity and $\chi_{\rm eff}$, where circular systems preferentially yield positive values while eccentric ones span a wider and potentially isotropic range.
    Furthermore, binaries that merge before achieving full alignment are expected to show enhanced 
    in-plane spin components ($\chi_P$) \citep{2025arXiv250608801F}. In such cases, even when the total spin magnitude is large, the effective spin can remain modest, offering a complementary signature of the AGN channel.
    This framework naturally encompasses massive events such as GW190521, with its observed spin–orbit tilt and possible eccentricity \citep{GW19a,2020ApJ...903L...5R,2022ApJ...940..171R,2022NatAs...6..344G}, and GW231123, whose highly spinning components may have grown through accretion \citep{2025arXiv250708219T}. Both could plausibly have originated in AGN disks \citep{2025arXiv250813412D,2025arXiv250808558B}. More broadly, AGN-assisted mergers may produce $\chi_{\rm eff}$ distributions ranging from strongly positive to nearly isotropic depending on merger properties, offering a direct observational test for future LVK analyses. This is consistent with current LVK results, which show $\chi_{\rm eff}$ values clustered around zero with a mild positive average-weighted distribution and only a small tail toward large spins \citep[e.g.][]{2023PhRvX..13a1048A}. A mixed population of mergers (circular, aligned binaries yield $\chi_{\rm eff}>0$ and eccentric, partially aligned systems remain near zero) in AGN disks can potentially explain this observed distribution. 

\end{itemize}

\subsection{Future Prospects}

The central result of this work is the emerging correlation between binary eccentricity and the effective spin parameter $\chi_{\rm eff}$ in AGN disks. 
Our analysis shows that quasi-circular binaries are likely to realign efficiently with the disk and therefore yield positive or near-zero $\chi_{\rm eff}$, 
while eccentric binaries can retain substantial spin–orbit misalignments unless alignment is extremely efficient. 
This establishes eccentricity as a potential predictor of the $\chi_{\rm eff}$ distribution in the AGN channel and provides a direct observational test for current and future gravitational wave detections.
Our framework represents a preliminary physical model for alignment in AGN disks. 
While not yet including the full complexity of binary–disk coupling, it captures the essential processes that determine whether binaries merge aligned or misaligned with the disk. 
A number of factors act to influence the effective alignment efficiency. Eccentric binaries shrink rapidly through GW emission, leaving less time for gas torques to act. Gas itself can accelerate orbital decay, as hydrodynamical simulations of binary–single encounters show that the compactness of triples decreases by up to two orders of magnitude in dense environments \citep[e.g.][]{Rowan2025_triples,Wang2025_triples,Wang2025_triplesII}.
This further limits alignment, especially in regions of the disk with high density and strong accretion.
Moreover, our present study has focused on BH–BH encounters, but the dominant interactions in AGN disks are likely between stars and stellar-mass BHs. 
Such star–BH scatterings can repeatedly perturb binaries during their inspiral, effectively altering the alignment process.

Advancing this work will require hydrodynamical simulations that couple the time evolution of $a_0$ and $e_0$ to tilt dynamics, as well as population studies that include both quasi-circular and eccentric binaries with distinct $\chi_{\rm eff}$ distributions. 
Such efforts are crucial for testing whether the AGN channel can explain the most puzzling events to date, such as GW190521 and GW231123, which combine high masses, potential eccentricity, and measurable spin–orbit tilt. Finally, we stress that the eccentricity–$\chi_{\rm eff}$ correlation identified here may not be unique to AGNs: analogous processes may operate in other dense, gas-rich environments such as nuclear stellar disks or star-forming regions \citep[e.g.][]{2025ApJ...983L...9K}. Determining whether this correlation is a signature specific to AGNs or a more universal feature of dynamical mergers is an exciting challenge for future theoretical and observational work.

\section*{Acknowledgments}
\label{sec:Acknowledgements}
We gratefully acknowledge valuable discussions with Philip Kirkeberg and David O'Neill, the participants of the DYNAMIX meeting in Cambridge, as well as Alex Dittmann, Barry McKernan and Saavik Ford, whose comments provided important insights for this work. 
The authors also thank the anonymous referee for the insightful suggestions
and comments.
The Center of Gravity is a Center of Excellence funded by the Danish National Research Foundation under grant No. 184.
The research leading to this work received funding from the Villum
Fonden Grant No. 29466, ERC Starting Grant no. 101043143,  the Independent Research Fund Denmark via grant ID 10.46540/3103-00205B, and by the European Union’s Horizon 2023 research and innovation program under Marie Sklodowska-Curie grant agreement No. 101148364.  The Tycho supercomputer hosted at the SCIENCE HPC center at the University of Copenhagen was used for performing the presented simulations.

\section*{Data Availability}
The data underlying this article will be shared on request to the corresponding author.

\bibliographystyle{mnras}
\bibliography{references}

\clearpage
\onecolumn
\appendix

\section{Disk Profile}

In this appendix we present the density and sound speed profiles of the \citetalias{Sirko03} model, shown as a function of distance from the SMBH in units of $R_{\rm S}$ and for different black-hole masses. These quantities are particularly relevant for evaluating the efficiency of binary alignment, since both the density $\rho_{\rm g}$ and sound speed $c_s$ directly enter the alignment timescale. The dashed contour highlights the transition between the accretion-dominated and tidally limited regimes, corresponding to the regions where gas accretion is either governed by the binary’s Bondi radius or limited by the Hill sphere. As discussed in Sec.~\ref{sec:prescription}, this distinction controls the scaling of $\dot{J}$ and therefore the efficiency of spin–orbit realignment.

\begin{figure*}
    \centering
    \includegraphics[width=\linewidth]{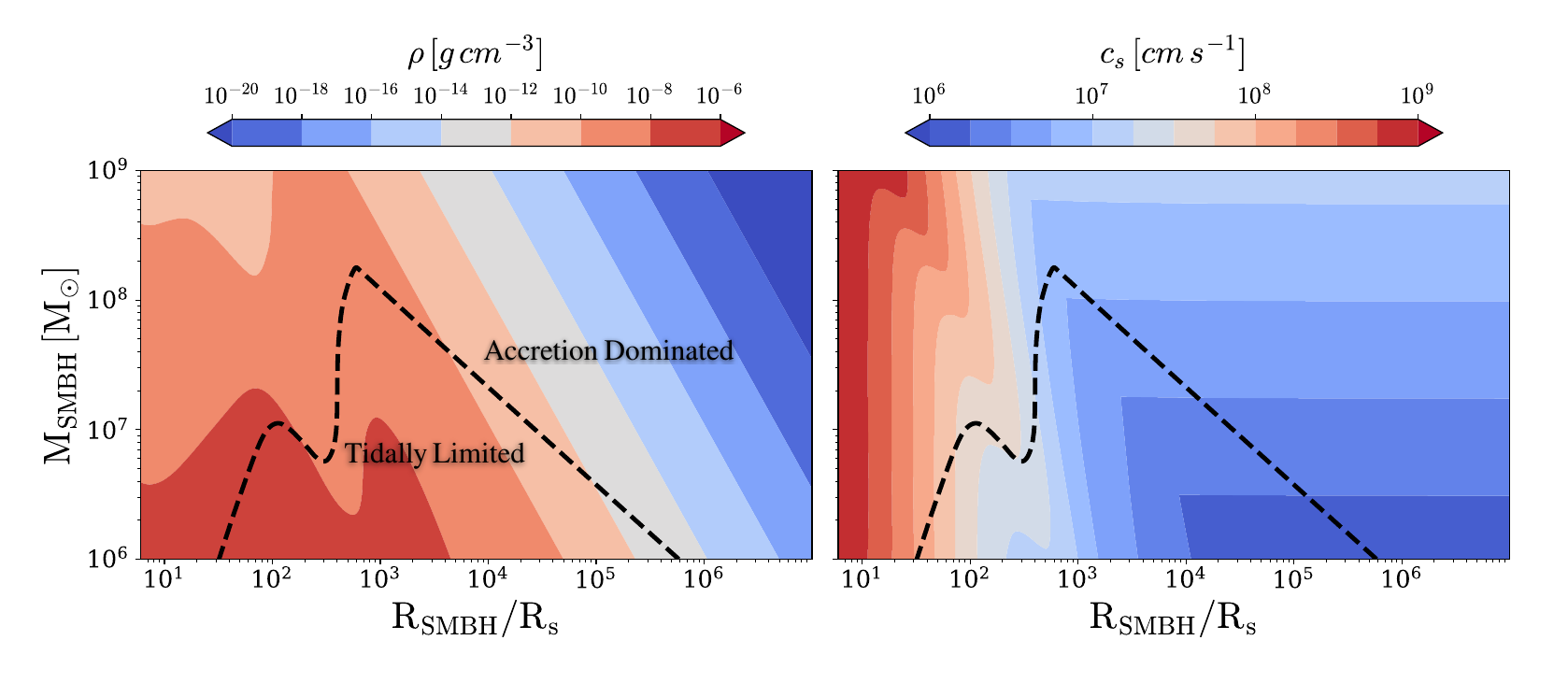}
    \caption{Density (left) and sound speed (right) profiles for the \citetalias{Sirko03} disk model, shown as a function of SMBH mass and distance from the SMBH in units of $R_{\rm S}$. The black dashed contour marks the boundary between the accretion-dominated (AD) and tidally limited (TL) regimes. Both $\rho_{\rm g}$ and $c_s$ are key inputs to the alignment prescription, with higher densities and lower sound speeds enhancing the efficiency of gas-driven realignment.}
    \label{fig:profiles}
\end{figure*}

\bsp
\label{lastpage}

\end{document}